\documentclass[aps,showpacs,amsmath,amssymb,superscriptaddress,footinbib,twocolumn]{revtex4}

\usepackage[english]{babel}

\usepackage{latexsym}
\usepackage{graphics}
\usepackage{epsfig}
\usepackage{amsfonts}
\usepackage{amssymb}
\usepackage{amsmath}

\begin{document}

\title{Dynamics of the Jaynes-Cummings and Rabi models: old wine in new bottles}

\author{Jonas Larson}

\affiliation{ICFO--Institut de Ci\`encies
Fot\`oniques, E-08860 Castelldefels, Barcelona, Spain}
\date{\today}

\begin{abstract}
By using a wave packet approach, this paper reviews the
Jaynes-Cummings model with and without the rotating wave
approximation in a non-standard way. This gives new insight, not
only of the two models themself, but of the rotating wave
approximation as well. Expressing the models by field quadrature
operators, instead of the typically used boson ladder operators,
wave packet simulations are presented. Several known phenomena of
these systems, such as collapse-rivivals, Rabi oscillation,
squeezing and entanglement, are reviewed and explained in this new
picture, either in an adiabatic or diabatic frame. The harmonic
shape of the potential curves that the wave packets evolve on and
the existance of a level crossing make these results interesting in
a broader sense than only for models in quantum optics, especially
in atomic and molecular physics.
\end{abstract}


\maketitle

\section{Introduction}\label{sec1}
Two fundamental models of quantum mechanics, presented in most
introductory textbooks, are the two-level system and the harmonic
oscillator. Combining these two into a bipartite system gives many
interesting models, where two of the more studied ones are the
Jaynes-Cummings (JC) model \cite{jc,jc2} and the Rabi model
\cite{rabi}. The JC model was introduced for describing the
interaction between a two-level atom and a quantized electromagnetic
(EM) field, while the Rabi model was introduced for NMR systems. In
this paper we will use the terminology of the JC model, a two-level
atom and a quantized EM field, which, of course, does not restrict
the results to such systems. The JC Hamiltonian is obtained from the
Rabi one by simply imposing the rotating wave approximation RWA
\cite{allen-eberly}. There are, however, special cases where atomic
selection rules make the JC model exact; the RWA terms naturally
vanish \cite{selrule}. In this approximation, exact analytical
solutions exist, and, in spite the simplicity of the JC model, the
dynamics have turned out to be very rich and complex, describing
several physical phenomena. Among these are; Rabi oscillations
\cite{rabi,jcrabi}, collapse-revivals \cite{cr}, squeezing
\cite{jcsqueez}, atom-field entanglement \cite{entangle},
non-classical states as Schr\"odinger cats \cite{cat} and Fock
states \cite{fock} and anti-bunching \cite{jcab}. The JC model
originally thought of as a single atom, single field mode
interaction, has shown to be applicable for several other types of
systems, for example, trapped ions \cite{ion}, Cooper-pair boxes
\cite{cooper}, "flux" qubits \cite{flux} and Josephson-junctions
\cite{jj}. With the experimental progress of some of the above
mentioned systems, the coupling between the systems may be made very
large, and the RWA breaks down so that only the Rabi model describes
the dynamics correctly. As the number of excitations, field plus
atom, is a constant of motion in the JC model, there are only two
physically relevant parameters; detuning $\Delta$ between the atomic
transition frequency $\Omega$ and the field mode frequency $\omega$
and atom-field coupling $g_0$. In the Rabi model, however, the
number of excitations is not conserved and all three parameters are
of importance. In general, the RWA is justified for small detunings
$\Delta=\Omega-\omega$ and small ratios of the atom-field coupling
divided by the atomic transition frequency, $g_0/\Omega$. In
atom-field cavity systems, this ratio is typically of the order
$g_0/\Omega\sim10^{-7}-10^{-6}$, \cite{csys}. Recently, cavity
systems with very strong couplings have been discussed
\cite{meystre}. The ratio may also become order of magnitudes larger
in solid state systems \cite{cooper}, and the full Rabi Hamiltonian,
including the {\it virtual processes} (also caller {\it
counter-rotating terms}), must be considered. The neglection of the
counter-rotating terms may have interesting physical consequencies
on nonlocality and causality \cite{nlc}. The effect of the RWA in
other systems has been studied, and to mention a few, trapping of
atoms by light \cite{trap}, driven Rabi model \cite{drivjc},
multi-level atoms and/or fields \cite{multi}, the micromaser
\cite{micro} semiclassical Rabi models \cite{semi}, transition
between vibrational states in a water molecule \cite{water}, dipolar
molecules \cite{drivmol}, light-matter interaction \cite{lightmat},
electromagnetically and self induced transparency \cite{eit,self}
and laser modified collisions \cite{lasercol}.

Existence of analytical solutions for special cases of the Rabi
model has been discussed \cite{rabiexsol}. As no general simple
solutions are known, much work has been devoted to various analytical approximate or numerical
approaches, such as; perturbation theory
\cite{pertur,jcrwa,rabiq,pegg,rabient,pertrabi}, path-integrals \cite{path},
continued fractions \cite{confra}, variational methods \cite{varm},
coupled cluster methods \cite{ccm}, displaced oscillators
\cite{displace,displace2}, approximated unitary transformed Rabi
Hamiltonians \cite{unitrans} and numerical studies
\cite{adwp,rabisqueez,disq,clsb,num46}. The time-dependent Rabi Hamiltonian has
been considered in \cite{timerabi}, and a particular equivalence
relation between the time-dependent JC and Rabi models was obtained.
Interestingly, the Rabi model has shown to be chaotic \cite{chaos},
contrary to the JC model \cite{jcchaos}. The Bargmann representation
\cite{barg}, similar to our approach,  has been considered for both
the JC \cite{stig} and the Rabi model \cite{rabibarg}.

Using a different approach than the standard ones we discuss and
review some of the work that has been done previously on the two
models. The method used in this paper is a numerical wave packet
propagation, which has been briefly applied to the Rabi model before
\cite{adwp,rabisqueez}, but to the best of our knowledge not to the
JC model. The system Hamiltonians are usually represented in the
field creation and annihilation operators $a^\dagger$ and $a$, but
here we instead work with the field quadrature operators $P$ and
$Q$. As these obey the standard canonical commutation relations, the
system is equivalent to a particle, with unit mass, moving in two
coupled potentials, and the dynamics is given by the evolution of
some initial wave packet. The idea of this paper is to give a deeper
understanding of the JC and Rabi models by consider the wave packets
evolving on these potential curves. Since the JC model is exactly
solvable, it might seem unmotivated to use this numerical study.
However, from the wave packet picture it is easy to understand some
of the systems behaviours, and, additionally, due to the
``graininess'' of the cavity field most analytically obtained
quantities predicted by the JC model are given as infinite sums
lacking closed forms. As in related problems in molecular physics
\cite{mol}, it is often convenient to investigate the system in the
{\it diabatic} or {\it adiabatic} basis
\cite{adwp,rabisqueez,clsb,adrabi}. The proper basis used, depends
on the system parameters and initial state. The critical evolution
takes place close to {\it level crossings} between the two energy
curves, and such a level crossing exists in both the JC and Rabi
model. The two energy curves are given, in both models, by two
displaced harmonic oscillators, and they are coupled by a constant
coupling in both cases, while in the JC model there is an additional
"momentum" $P$ dependent coupling which gives rise to very different
dynamics. The presence of the $P$ dependent coupling makes the wave
packet evolution of the JC model less intuitive and more complex.
Comparing the two models, combined with the knowledge of the
validity regime of the RWA, one may interpret the consequences of
this coupling term as well as other effects. In the intermediate
regime, when one is neither in the adiabatic nor the diabatic
validity regimes, a more detailed analysis is needed to understand
the full evolution. This is only briefly mentioned in this paper.

The paper proceeds as the following. In the next section \ref{sec2}
we introduce the JC and Rabi Hamiltonians in their regular forms and
also give them in the conjugate variable picture. In subsection
\ref{sslz} the corresponding curve crossing problem is mentioned and
how to approximate it as a Landau-Zener problem \cite{lz}. The
following section \ref{sec3}, presents a background of an adiabatic
approach, the two different bases, diabatic and adiabatic, are
defined, and the approach is applied to both the JC and Rabi model.
The validity of the adiabatic method is numerically studied in Sec.
\ref{sec4} using the split operator procedure to calculate the
fidelity. In Sec. \ref{sec5} we present the numerical results of our
studies. Several plots are shown to give a deeper insight of the
dynamics and the phenomena are discussed from a wave packet point of
view. Known interesting effects, such as entanglement,
collapse-revivals, squeezing and Rabi oscillations, are especially
studied. Finally we conclude the paper in sec. \ref{conc} with a
summery.

\section{The Jaynes-Cummings and Rabi models}\label{sec2}
\subsection{Introducing the JC model with and without RWA}
The simplest fully quantum mechanical model describing light-matter
interaction, naturally considers one atom, with a single atomic
transition, interacting with only one mode of the field. Such an
approach leads to the JC or Rabi models. The single atom transition
and single mode approximation relies on small or vanishing coupling
elements between other levels \cite{scully}. Thus, mathematically
one is left with a two-level atom (spin-1/2 particle) coupled to a
harmonic oscillator. For a dipole interaction,
$H_{int}=\bar{d}\cdot\bar{E}/\hbar$, where $\bar{d}$ and $\bar{E}$
are the atomic dipole moment and electric field respectively, a
microscopic derivation \cite{scully} gives the Rabi Hamiltonian
\begin{equation}\label{hrabi}
H_{Rabi}=\hbar\omega \left(a^\dagger
a+\frac{1}{2}\right)+\frac{\hbar\Omega}{2}\sigma_z+\hbar
g_0\left(\sigma^++\sigma^-\right)\left(a^\dagger+a\right).
\end{equation}
Here, $a^\dagger$ ($a$) is the creation (annihilation) operator for the
field mode; $a|n\rangle=\sqrt{n+1}|n+1\rangle$
($a|n\rangle=\sqrt{n}|n-1\rangle$), the sigma operators are the
standard Pauli matrices acting on the two-level atom;
$\sigma_z|\pm\rangle=\pm|\pm\rangle$,
$\sigma^\pm|\pm\rangle=|\mp\rangle$, $\sigma_x=\sigma^++\sigma^-$
and $\sigma_y=-i\left(\sigma^+-\sigma^-\right)$, $\omega$ ($\Omega$)
is the field (atomic transition) frequency and $g_0$ is the
effective atom-field coupling. In deriving (\ref{hrabi}), the dipole
approximation has been assumed, where the variation of the field is
neglected on the atomic length scale, and the kinetic energy of the
atom is omitted, valid for large and moderate temperatures \cite{effmass}.

The interaction part contains four terms; $\sigma^+a^\dagger$ ($\sigma^-a$) simultaneous
excitation (de-excitation) of the atom and field, and $\sigma^+a$
($\sigma^-a^\dagger$) excitation of the atom by absorption of one
photon (de-excitation of the atom by emission of one photon). In the
Heisenberg picture, the operators evolve (free field and free
atom case) as: $a(t)=a\exp(i\omega t)$,
$a^\dagger(t)=a^\dagger\exp(-i\omega t)$ and
$\sigma^\pm(t)=\sigma^\pm\exp(\pm i\Omega t)$. Thus, the interaction
terms precess with either the frequencies $|\Omega-\omega|$ (energy
conserving terms) or $\Omega+\omega$ (non energy conserving/counter rotating terms).
Most often $|\Omega-\omega|\ll\Omega+\omega$ and the fast
oscillating terms are rejected from the Hamiltonian, resulting in
the Jaynes-Cummings Hamiltonian
\begin{equation}\label{hjc0}
H_{JC}=\hbar\omega \left(a^\dagger
a+\frac{1}{2}\right)+\frac{\hbar\Omega}{2}\sigma_z+\hbar
g_0\left(\sigma^+a+\sigma^-a^\dagger\right).
\end{equation}
However, for a large atom-field detuing (but still within the atomic
two-level and single mode approximations), the above constrain may
be violated and the non energy conserving terms can not be excluded.
As mentioned in the introduction, also the ratio $g_0/\Omega$ is
important for the validity of the above RWA. The two conditions for
the RWA must in general be met simultaneously, meaning that near
resonant interaction does not gurantee validity of the RWA if not
$g_0<\Omega$, and vice versa.

\subsection{The models in the conjugate variable picture}
The JC model is usually given and solved within the "number"-basis
of the field $\{|n\rangle|\pm\rangle\}$. Here, however, we will work
in the conjugate variable basis of the field. We introduce the
conjugate variables $P$ and $Q$ related to the creation/annihilation
operators as
\begin{equation}
\begin{array}{l}
P=i\sqrt{\frac{\hbar\omega}{2}}\left(a^\dagger-a\right), \\ \\
Q=\sqrt{\frac{\hbar}{2\omega}}\left(a^\dagger+a\right).
\end{array}
\end{equation}
In the transformed basis the Rabi (\ref{hrabi}) and the JC
(\ref{hjc0}) Hamiltonians read
\begin{equation}\label{conrabi}
\displaystyle{H_{Rabi}=\frac{P^2}{2}+\frac{\omega^2}{2}Q^2+\left[\begin{array}{cc}\displaystyle{\frac{\hbar\Omega}{2}}
& g_0\sqrt{2\hbar\omega}Q \\
g_0\sqrt{2\hbar\omega}Q &
-\displaystyle{\frac{\hbar\Omega}{2}}\end{array}\right]},
\end{equation}

\begin{equation}\label{conjc}
\begin{array}{c}
\displaystyle{H_{JC}=\frac{P^2}{2}+\frac{\omega^2}{2}Q^2}
\\ \\ \displaystyle{+\left[\begin{array}{cc}\displaystyle{\frac{\hbar\Omega}{2}} &
\displaystyle{g_0\sqrt{\frac{\hbar}{2}}\left(\sqrt{\omega}Q+i\frac{P}{\sqrt{\omega}}\right)}
\\
\displaystyle{g_0\sqrt{\frac{\hbar}{2}}\left(\sqrt{\omega}Q-i\hbar\frac{P}{\sqrt{\omega}}\right)}
& -\displaystyle{\frac{\hbar\Omega}{2}}\end{array}\right]}.
\end{array}
\end{equation}
Relating $P$ and $Q$ as momentum and position operators
respectively, the above Hamiltonians can be interpreted as a
two-level particle, with quantized position and momentum, confined
in a harmonic trap and interacting with a "classical" field with
mode "variations" $g(Q,P)\sim Q$ or
$g(Q,P)\sim\left(Q+i\frac{P}{\omega}\right)$. The Rabi Hamiltonian
could be obtained, for example, by considering a harmonically
trapped two-level atom driven by a classical field, whose wave
length is much longer than the extent of the atomic wave packet in
the trap and with a node at the centre $Q=0$. In other words, there
is a direct connection between the above example of a trapped atom
and an atom interacting with a quantized field. Relating the JC
Hamiltonian to similar systems is less trivial, since the atom-field
coupling is momentum dependent.

\subsection{The corresponding curve crossing problem}\label{sslz}
The structure of the Rabi Hamiltonian (\ref{conrabi}) is more easily
seen by applying the unitary transformation
$U=\frac{1}{\sqrt{2}}\left(\sigma_x+\sigma_z\right)$, giving the
transformed Hamiltonian
\begin{equation}
\begin{array}{c}
\displaystyle{\tilde{H}_{Rabi}=U^\dagger
H_{Rabi}U=\frac{P^2}{2}+\frac{\omega^2}{2}Q^2}
\\ \\
+\left[\begin{array}{cc}g_0\sqrt{2\hbar\omega}Q
&  \displaystyle{\frac{\hbar\Omega}{2}} \\
\displaystyle{\frac{\hbar\Omega}{2}} &
-g_0\sqrt{2\hbar\omega}Q\end{array}\right].
\end{array}
\end{equation}
By completing the squares we may write the Hamiltonian as
\begin{equation}
\tilde{H}_{Rabi}\!=\!\frac{P^2}{2}+\left[\!\begin{array}{cc}V_h\left(\!Q\!+\!\sqrt{\frac{2\hbar}{\omega^3}}g_0\!\!\right)
&  \displaystyle{\frac{\hbar\Omega}{2}} \\
\displaystyle{\frac{\hbar\Omega}{2}} &
V_h\left(\!Q\!-\!\sqrt{\frac{2\hbar}{\omega^3}}g_0\!\!\right)\end{array}\!\right]-\frac{
\hbar g_0^2}{\omega},
\end{equation}
where $V_h(x)=\omega^2x^2/2$ is the regular harmonic oscillator
potential. Thus, it is seen that the problem is equivalent to the
one of a particle moving in two coupled equally, but opposite,
displaced harmonic potential curves. The two energy curves cross for
$Q=0$, and the off diagonal coupling $\hbar\Omega/2$ results in an
avoided crossing. As is well known; the main population
transfer between two levels occurs close to the crossing. Here the distance between
the curves relative to the coupling amplitude is the smallest. Therefor, in some situations the
coupled dynamics may be considered only in the vicinity of the
crossing. By linearizing the model around $Q=0$ we get
\begin{equation}
\tilde{H}_{Rabi}=\frac{P^2}{2}+\left[\begin{array}{cc}
g_0\sqrt{2\hbar\omega}Q
&  \displaystyle{\frac{\hbar\Omega}{2}} \\
\displaystyle{\frac{\hbar\Omega}{2}} & -g_0\sqrt{2\hbar\omega}Q
\end{array}\right].
\end{equation}
If the wave packet is well localized (the
characteristic width of the wave packet $\Delta_Q$ is small compared
to variations of the harmonic potential), the "momentum" and
"position" may be replaced by its classical counterparts at the
crossing; $P\rightarrow v$ and $Q\rightarrow vT$, where $T$ is the
time. Neglecting the constant terms, the Hamiltonian reduces to the one of the
Landau-Zener model \cite{lz}
\begin{equation}\label{lzham}
\tilde{H}_{Rabi}=\left[\begin{array}{cc}
\sqrt{2\hbar\omega}g_0vT &
\displaystyle{\frac{\hbar\Omega}{2}} \\
\displaystyle{\frac{\hbar\Omega}{2}} &
-\sqrt{2\hbar\omega}g_0vT\end{array}\right],
\end{equation}
where the classical velocity $v$ will, of course, depend on the
initial state of the system. The time dependent problem given by the
Hamiltonian (\ref{lzham}) is analytically solvable, with asymptotic
solution for the population transfer to the other level
given by
\begin{equation}
P_{LZ}=1-\exp\left(-\frac{\pi\Omega^2}{4\sqrt{2\hbar\omega}g_0v}\right).
\end{equation}
The validity of this semiclassical model may be studied in a
detailed way similar to the one in \cite{holthaus}.

The same procedure may be applied to the JC Hamiltonian
(\ref{conjc});
\begin{equation}
\tilde{H}_{JC}\!=\!\frac{P^2}{2}\!+\!\left[\!\begin{array}{cc}V_h\left(Q\!+\!\sqrt{\frac{\hbar}{2\omega^3}}g_0\!\!\right)
&  \displaystyle{\frac{\hbar\Omega}{2}\!-\!ig_0\sqrt{\frac{\hbar}{2\omega}}P} \\
\displaystyle{\frac{\hbar\Omega}{2}\!+\!ig_0\sqrt{\frac{\hbar}{2\omega}}P}
&
V_h\!\left(\!Q\!-\!\sqrt{\frac{\hbar}{2\omega^3}}g_0\!\!\right)\end{array}\!\right]\!-\!\frac{
\hbar g_0^2}{4\omega}.
\end{equation}
One notes that the displaced oscillators are
shifted half the amount as in the Rabi Hamiltonian, and that there
is an additional $P$-dependent coupling of the oscillators. As the RWA implies that non energy conserving terms are omitted, it
follows that the number of excitations is conserved. Thus,
$N=\frac{P^2}{2\omega}+\frac{\omega}{2}Q^2+\frac{1}{2}\sigma_z$
commutes with $H_{JC}$, and one may work in a rotating frame with respect to $N$. Applying the same unitary operator to the new interaction picture Hamiltonian $H'_{JC}=H_{JC}-\hbar\omega N$, one finds
\begin{equation}\label{jclz}
\tilde{H}'_{JC}=\left[\begin{array}{cc} \displaystyle{g_0\sqrt{\frac{\hbar\omega}{2}}Q} & \displaystyle{\frac{\hbar\Delta}{2}-ig_0\sqrt{\frac{\hbar}{2\omega}}P} \\ \\
                       \displaystyle{\frac{\hbar\Delta}{2}+ig_0\sqrt{\frac{\hbar}{2\omega}}P} & \displaystyle{-g_0\sqrt{\frac{\hbar\omega}{2}}Q}
                      \end{array}\right]
\end{equation}
The energy transfer between the atom and the field in the JC model
is give by $\tilde{H}'_{JC}$, and from (\ref{jclz}) we note the
similarities with the Landau-Zener problem but with the addition
``momentum'' dependent coupling. An interesting observation is that
if the wave packet approaches the curve crossing with a high
momentum, an adiabatic evolution is still possible due to the large
curve couplings. At the one hand, ``velocity'' determines the
steepness of the crossing curves, but on the other is also
determines level separation close to the coupling. This peculiar
curve crossing model is, of course, an artifact of the RWA, but none
the less it may give insight into other related systems, for example
atomic and molecular scattering processes \cite{ccross}.

\section{The adiabatic approach}\label{sec3}
\subsection{Review of the adiabatic principle}
The concept of adiabaticity is widely used in a variety of physical
and chemical systems, while, what is known as the adiabatic theorem
\cite{adth} is {\it only} stated for time-dependent Hamiltonian
systems. Non the less, the principle may be transferred to other
systems where the dynamics is governed by the evolution of some
initial state \cite{adgen}. Here we review the results of \cite{jonas1}, and
apply it to our models in the following subsections.

Given a Hamiltonian of the form
\begin{equation}\label{adh1}
H=\frac{p^2}{2m}+V(x)+\left[\begin{array}{cc}\displaystyle{\frac{\hbar\Delta(x,p)}{2}}
& \hbar g(x,p) \\ \hbar g(x,p) &
-\displaystyle{\frac{\hbar\Delta(x,p)}{2}}\end{array}\right].
\end{equation}
The diagonal elements are often refereed to as detuning, and for
simplicity they will be taken constant $\Delta(x,p)=\Delta$, and the
off diagonal ones are level couplings. We introduce the parameter
$\theta$ accordingly
\begin{equation}\label{vinkel}
\tan2\theta=\frac{2g(x,p)}{\Delta},
\end{equation}
such that the unitary transformation
\begin{equation}
U=\left[\begin{array}{cc}\cos\theta & -\sin\theta \\ \sin\theta &
\cos\theta\end{array}\right]
\end{equation}
diagonalizes the last term of the Hamiltonian (\ref{adh1}) with
corresponding eigenvalues
\begin{equation}
\lambda_\pm=\pm\lambda=\pm\hbar\sqrt{\left(\frac{\Delta}{2}\right)^2+g^2(x,p)},
\end{equation}
and eigenstates, defining the new {\it adiabatic basis states},
\begin{equation}\label{adbas}
|\uparrow\rangle=\left[\begin{array}{c}\cos\theta \\ \sin\theta
\end{array}\right], \hspace{1cm}
|\downarrow\rangle=\left[\begin{array}{c}-\sin\theta \\ \cos\theta
\end{array}\right].
\end{equation}
However, since in general $[p,g(x,p)]\neq0$ and
$[V(x),g(x,p)]\neq0$, the new Hamiltonian will not be diagonal.
Introducing the two functions $f=\cos\theta$ and $h=\sin\theta$, and
using the identity
\begin{equation}
U^\dagger pU=p-\sigma_y\hbar\partial\theta,
\end{equation}
where
\begin{equation}
\partial\theta\equiv\frac{\partial\theta}{\partial x}
\end{equation}
the transformed Hamiltonian takes the form
\begin{equation}\label{transham}
\tilde{H}=\tilde{H}_{ad}+\tilde{H}_{cor},
\end{equation}
with the adiabatic part
\begin{equation}
\tilde{H}_{ad}=\frac{p^2}{2m}+(\hbar\partial\theta)^2+fV(x)f+hV(x)h+\left[\begin{array}{cc}\lambda
& 0 \\ 0 & -\lambda\end{array}\right]
\end{equation}
and the correction part
\begin{equation}
\begin{array}{lll}
\tilde{H}_{cor} & = & \frac{1}{2m}\left[\begin{array}{cc}0 &
\hbar^2\partial^2\theta-2i\hbar(\partial\theta)p \\
-\hbar^2\partial^2\theta+2i\hbar(\partial\theta)p &
0\end{array}\right] \\ \\
& & +\left(fV(x)h-hV(x)f\right)i\sigma_y.
\end{array}
\end{equation}
Note that if $V(x)$ commutes with $g(x,p)$ we have
$fV(x)f+hV(x)h=V(x)$ and $fV(x)h-hV(x)f=0$. As it will turn out,
this will be the case in our models, and we therefor
assume commutability.

An adiabatic evolution is obtained if the dynamics is dominated by
the diagonal adiabatic Hamiltonian $H_{ad}$ of (\ref{transham}),
while $H_{cor}$ only slightly affects the propagation. Clearly, this
depends on the smoothness of $\theta$ and the momentum $p$, but also
on the wave packet shape. Using the definition of the angle
parameter (\ref{vinkel}) we find
\begin{equation}
\begin{array}{l}
\displaystyle{\partial\theta=\frac{\Delta\partial g}{\Delta^2+4g^2}}, \\ \\
\displaystyle{\partial^2\theta=\frac{\Delta\left(\partial^2
g\left(\Delta^2+4g^2\right)-8g(\partial
g)^2\right)}{\left(\Delta^2+4g^2\right)^2}}.
\end{array}
\end{equation}
Under the adiabatic approximation, the state is evolved on the {\it
adiabatic energy curves} $\varepsilon_\pm=V\pm\lambda$ according to
\begin{equation}\label{adinstate}
\begin{array}{lll}
\Psi^{ad}(x,t) & = &
\exp\left[-i\left(\frac{p^2}{2m}+\varepsilon_+\right)t\right]\psi_\uparrow^{ad}(x,0)|\uparrow\rangle\\
\\
& & +
\exp\left[-i\left(\frac{p^2}{2m}-\varepsilon_-\right)t\right]\psi_\downarrow^{ad}(x,0)|\downarrow\rangle.
\end{array}
\end{equation}
In the opposite limit, the non-adiabatic case, the evolution takes
place on the {\it diabatic energy curves} (also called bare
energies), defined by the diagonal elements of the Hamiltonian. Note,
though, that the diabatic energies depend on the basis used.

\subsection{Application to the JC model}
Before proceeding, we introduce scaled dimensionless
variables. We scale energies with the photon energy $\hbar\omega$,
"lengths" $Q$ by $\sqrt{\hbar/\omega}$ and time by $\omega^{-1}$:
\begin{equation}
\begin{array}{lll}
\displaystyle{q=\sqrt{\frac{\omega}{\hbar}}Q}, & & t=\omega T \\ \\
\displaystyle{\tilde{\Omega}=\frac{\Omega}{\omega}}, & &
\displaystyle{\tilde{g}_0=\frac{g_0}{\omega}}.
\end{array}
\end{equation}
For convenience we drop the tildes $\sim$ on the last two scaled
variables.

Introduced in the subsection \ref{sslz}, it is convenient to work in an interaction picture with
respect to excitation operator,
\begin{equation}
H'_{JC}=H_{JC}-N=\left[\begin{array}{cc}
\displaystyle{\frac{\Delta}{2}} &
\displaystyle{g_0\frac{1}{\sqrt{2}}\left(q+ip\right)}
\\ \\
\displaystyle{g_0\frac{1}{\sqrt{2}}\left(q-ip\right)} &
\displaystyle{-\frac{\Delta}{2}}\end{array}\right],
\end{equation}
with the detuning $\Delta=\Omega-1$. As the interaction picture
Hamiltonian is lacking the kinetic energy operator and the "external
potential", the correction $H_{cor}$ to the adiabatic Hamiltonian
vanishes. The eigenstates are
\begin{equation}\label{jcsol}
\begin{array}{l}
|\uparrow\rangle_n=\cos\theta\psi_{n-1}(q)|+\rangle+\sin\theta\psi_{n}(q)|-\rangle,
\\ \\
|\downarrow\rangle_n=\sin\theta\psi_{n-1}(q)|+\rangle-\cos\theta\psi_{n}(q)|-\rangle,
\end{array}
\end{equation}
where
\begin{equation}
\tan2\theta=\frac{2g_0\sqrt{n}}{\Delta},
\end{equation}
and $\psi_n(q)$ is the $n$th eigenfunctions to the harmonic
oscillator. As $q\pm ip$ act as raising and lowering operators, the
diagonaliced Hamiltonian in the original picture and within the
$n$th block reads
\begin{equation}
H_{JC}=\frac{p^2}{2}+\frac{q^2}{2}+\frac{1}{2}\sigma_z\pm\sqrt{\left(\frac{\Delta}{2}\right)^2+g_0^2n}.
\end{equation}
Thus, the eigenvalues are
\begin{equation}\label{jcsol2}
\displaystyle{E_\pm(n)=\left(n+\frac{1}{2}\right)\pm\sqrt{\left(\frac{\Delta}{2}\right)^2+g_0^2n}}.
\end{equation}

The "adiabatic limit" in the JC model is generally taken as
$\Delta\gg g_0\sqrt{n}$. It has been shown \cite{jonas1} that this
limit governs adiabatic evolution, but that this is not the only
possibility. Expanding the solutions (\ref{jcsol}) and
(\ref{jcsol2}) to first order in $g_0\sqrt{n}/\Delta$ reproduce the
familiar results from "adiabatic elimination" in the JC model
\cite{adel}.

\subsection{Application to the Rabi model}
Applying the above method to the Rabi Hamiltonian (\ref{conrabi})
give the adiabatic Hamiltonian
\begin{equation}\label{adh2}
H_{ad}=\frac{p^2}{2}+V_\pm(q),
\end{equation}
where
\begin{equation}\label{adpot}
V_\pm(q)=\frac{q^2}{2}+\frac{2\Omega^2g_0^2}{\left(\Omega^2+2g_0^2q^2\right)^2}\pm\sqrt{\left(\frac{\Omega}{2}\right)^2+2g_0^2
q^2}
\end{equation}
and the corresponding adiabatic states given in (\ref{adbas}). In
the limit $\Omega\rightarrow0$, the adiabatic correction vanishes
and the problem is analytically solvable, since (\ref{adh2}) becomes
identical to two displaced disconnected harmonic oscillators.

The standard way of defining an adiabaticity criteria is by letting
the "distance" between the adiabatic energies become much larger than
the amplitude of the off-diagonal couplings \cite{adth}. In our
particular model, we explicitly have
\begin{equation}
\begin{array}{l}
\displaystyle{\partial\theta=\frac{\Omega
g_0\sqrt{2}}{\Omega^2+2g_0^2q^2}}, \\ \\
\displaystyle{\partial^2\theta=-\frac{4\Omega
g_0^3\sqrt{2}q}{\left(\Omega^2+2g_0^2q^2\right)^2}}.
\end{array}
\end{equation}
For smooth angles $\theta$, the main non-adiabatic contribution will
come from the term $2(\partial\theta)p$, where a large $p$ is
obtained by having a large $\langle n\rangle$ which is known to
violate adiabaticity. We directly note that the standard adiabatic
limit $\Omega\gg g_0$ gives a small coupling. However, we also note
that if $q^2\gg\left(\sqrt{2}g_0\Omega-\Omega^2\right)/2g_0^2$, an
adiabatic evolution is obtained, which is intuitive as we are far
from the curve crossing. It is instructive to analyze the diabatic
and adiabatic energy curves for different sets of parameters which
is shown in fig. 1. Note that the diabatic energies are just the
displaced harmonic oscillators $V_h(q\pm\sqrt{2}g_0)$. In (a), the
atomic frequency $\Omega$ is larger than the coupling and we have in
this situation the regular adiabatic limit. In (b)-(d) $\Omega$ is
smaller or equal to $g_0$ and an adiabatic evolution is only
expected when the wave packet is well localized at positions
$|q|\gg0$. We note how the barrier in the middle at $q=0$ of the
adiabatic curves is lower, but wider, for increasing $\Omega$. This
peak will most likely govern a non-adiabatic transition between the
adiabatic states. This again, indicates that adiabatic evolution is
only obtained far from the crossing in these situations. For large
enough $|q|$ the adiabatic and diabatic states coincide for any
parameters. When $\Omega/g_0$ is small, a large $|q|$ is needed to
assure adiabaticity, while a large $\Omega/g_0$ induces adiabaticity
for all "positions" $q$. Thus, since the states we propagate will
most often pass through the level crossing we conclude that for
$\Omega>g_0$ the behaviour may be understood from the evolution on
the adiabatic energy curves, while for $\Omega<g_0$ it is better to
use the diabtic curves.

For simplicity, the initial state will be given by the atom in its
excited state and the field in some photon distribution, hence
\begin{equation}\label{instate}
\Psi(q,0)=\psi_+(q,0)|+\rangle=\psi_+(q,0)\left(\cos\theta|\uparrow\rangle-\sin\theta|\downarrow\rangle\right).
\end{equation}
This limits the analysis, since more effects may be seen with other
initial conditions, but that would greatly extend the leangth of the paper. For an initial Fock state, $a^\dagger
a|n\rangle=n|n\rangle$, or coherent state,
$a|\nu\rangle=\nu|\nu\rangle$, we have in $q$-representation
respectively
\begin{equation}
\begin{array}{l}
\displaystyle{\psi_n(q,0)=\frac{1}{\sqrt{2^nn!}}\left(\frac{1}{\pi}\right)^{1/4}\mathrm{H}_n(q)\mathrm{e}^{-\frac{q^2}{2}}},\\
\\
\displaystyle{\psi_\nu(q,0)=\left(\frac{1}{\pi}\right)^{1/4}\mathrm{e}^{-(\Im\nu)^2}\mathrm{e}^{-\frac{1}{2}\left(q-\sqrt{2}\nu\right)^2}},
\end{array}
\end{equation}
where $\mathrm{H}_n$ is the $n$th order Hermite polynomial. We note
that for initial Fock states, adiabaticity is not expected if
$\Omega<g_0$, since their wave functions are non-zero around $q=0$.
However, for odd $n$ we have $\psi_n(q=0,0)=0$, but numerics
indicate that adiabaticity is lost also for such states.

\begin{figure}[ht]
\begin{center}
\includegraphics[width=8cm]{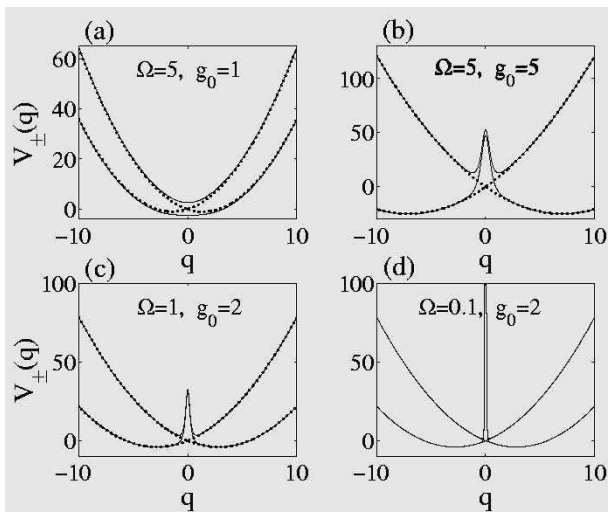}
\caption{\label{fig1} Solid lines display the adiabatic energy
curves (\ref{adpot}) for four different sets of dimensionless
parameters indicated in the plots. The dotted lines show the
corresponding displaced harmonic potentials $V_h(q\pm\sqrt{2}g_0)$. Here the "momentum" $p=1$.}
\end{center}
\end{figure}

\section{Validity check of the adiabatic approximation}\label{sec4}
The validity of the adiabatic approximation is studied from a full
wave packet simulation. The method used is the split operator
procedure \cite{split}, which relies on separation of the evolution
operator into a momentum and position dependent one. The wave packet
is either given in the $q$- or the $p$-representation, by a simple
fast Fourier transform (FFT). The initial state of the field is
given in eq. (\ref{instate}), and it will be propagated by, either
the adiabatic Hamiltonian (\ref{adh2}) or the full Hamiltonian
(\ref{conrabi}). From the corresponding states, labelled
$\Psi^{ad}(q,t)$ and $\Psi(q,t)$ respectively, we measure the
validity of the approximation with the fidelity
\begin{equation}\label{fidel}
F(t)=\sqrt{\left|\int\Psi^*(q,t)\Psi^{ad}(q,t)\,dq\right|}.
\end{equation}
Numerical simulations indicate that a large detuning, $\Omega\ll1$
or $\Omega\gg1$, governs an adiabatic evolution. These limits are
opposite to the one related to validity of the RWA. However, the
limit $(g_0/\Omega)\rightarrow0$ assures applicability of the
adiabatic approximation and RWA. Non the less, there are parameter
regimes where the adiabatic approximation is justified, while RWA
breaks. Particularly, for a fixed ratio $g_0/\Omega$, the limits
$\Omega\rightarrow\infty$ or $\Omega\rightarrow0$ guarantee
adiabaticity.

\begin{figure}[ht]
\begin{center}
\includegraphics[width=8cm]{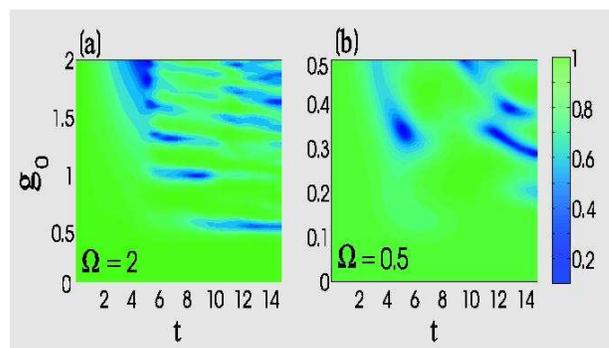}
\caption{\label{fig2} (Colour online)Plots of the fidelity (\ref{fidel}), for an
initial excited atom and empty cavity, as a function of scaled time
$t$ and atom-field coupling $g_0$. In (a) the atom transition
frequency is $\Omega=2$ and in (b) $\Omega=0.5$. The colourbar to the
right displays the value of $F(t)$. }
\end{center}
\end{figure}

\begin{figure}[ht]
\begin{center}
\includegraphics[width=8cm]{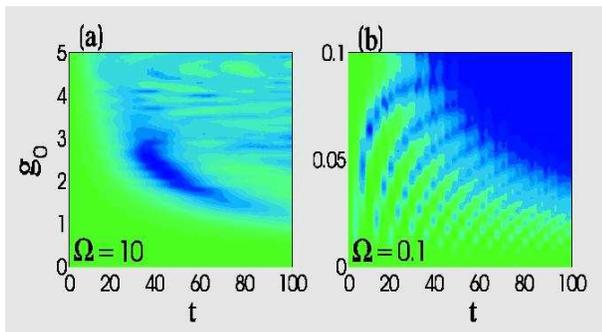}
\caption{\label{fig3} (Colour online) The same as fig. \ref{fig2}, but for an
initial coherent state with $\langle n\rangle=|\nu|^2=16$. The
atomic transition frequencies are $\Omega=10$ and $\Omega=0.1$. The
colour scale is the same as in fig. \ref{fig2}. }
\end{center}
\end{figure}

In fig. \ref{fig2} we show the dependence of the fidelity
(\ref{fidel}), for an initial empty cavity (Fock state), on the
coupling amplitude $g_0$ and time $t$. In the left plot $\Omega=2$
and in the right one $\Omega=0.5$, and it is clear that keeping
$g_0/\Omega$ small results in an adiabatic evolution. The next plot,
fig \ref{fig3}, gives the same, but now for an initial coherent
state of the cavity field with an amplitude $\nu=4$. In (a)
$\Omega=10$ and in (b) $\Omega=0.1$. The colour scale is the same
for both figures. By lowering the amplitude $\nu$ the fidelity would
be increased. It should be emphasised again that for a larger
detuning between the mode and atomic transition frequency, one would
also obtain a more adiabatic evolution.

\section{Numerical investigations}\label{sec5}
In this section we investigate in a non-standard way some of the
more well-known phenomena predicted by the JC model. Wave packet
propagation will be presented for both the JC and Rabi model. As
mentioned in the introduction, one advantage of the method of using
wave packet propagation is that it is valid for all parameter
regimes. However, one should keep in mind that the adiabatic
approximation is only valid within certain parameter regimes as seen
in the previous section. The numerical results will be analyzed in
terms of the wave packets propagating on the appropriate adiabatic
or diabatic energy curves.

\subsection{Rabi oscillations}

\begin{figure}[ht]
\begin{center}
\includegraphics[width=8cm]{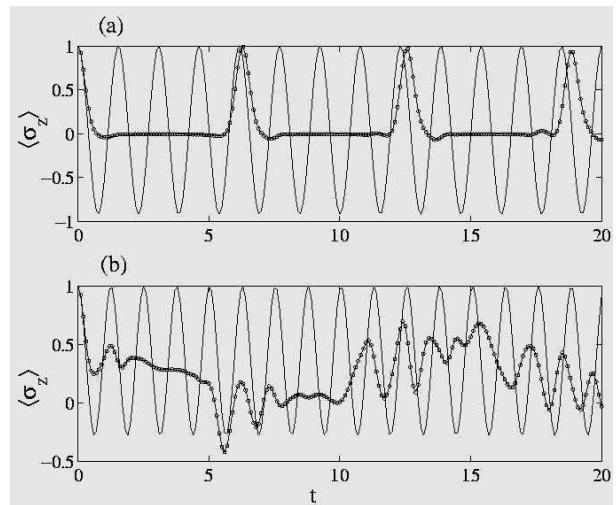}
\caption{\label{fig4} The figure shows the atomic inversion (Rabi
oscillations) for an initial empty cavity and excited atom. Circles
give the non-RWA results, while solid lines the RWA ones. In (a)
$\Omega=0.2$ and $g_0=2$ and in (b) $\Omega=4$ and $g_0=2$. }
\end{center}
\end{figure}

\begin{figure}[ht]
\begin{center}
\includegraphics[width=8cm]{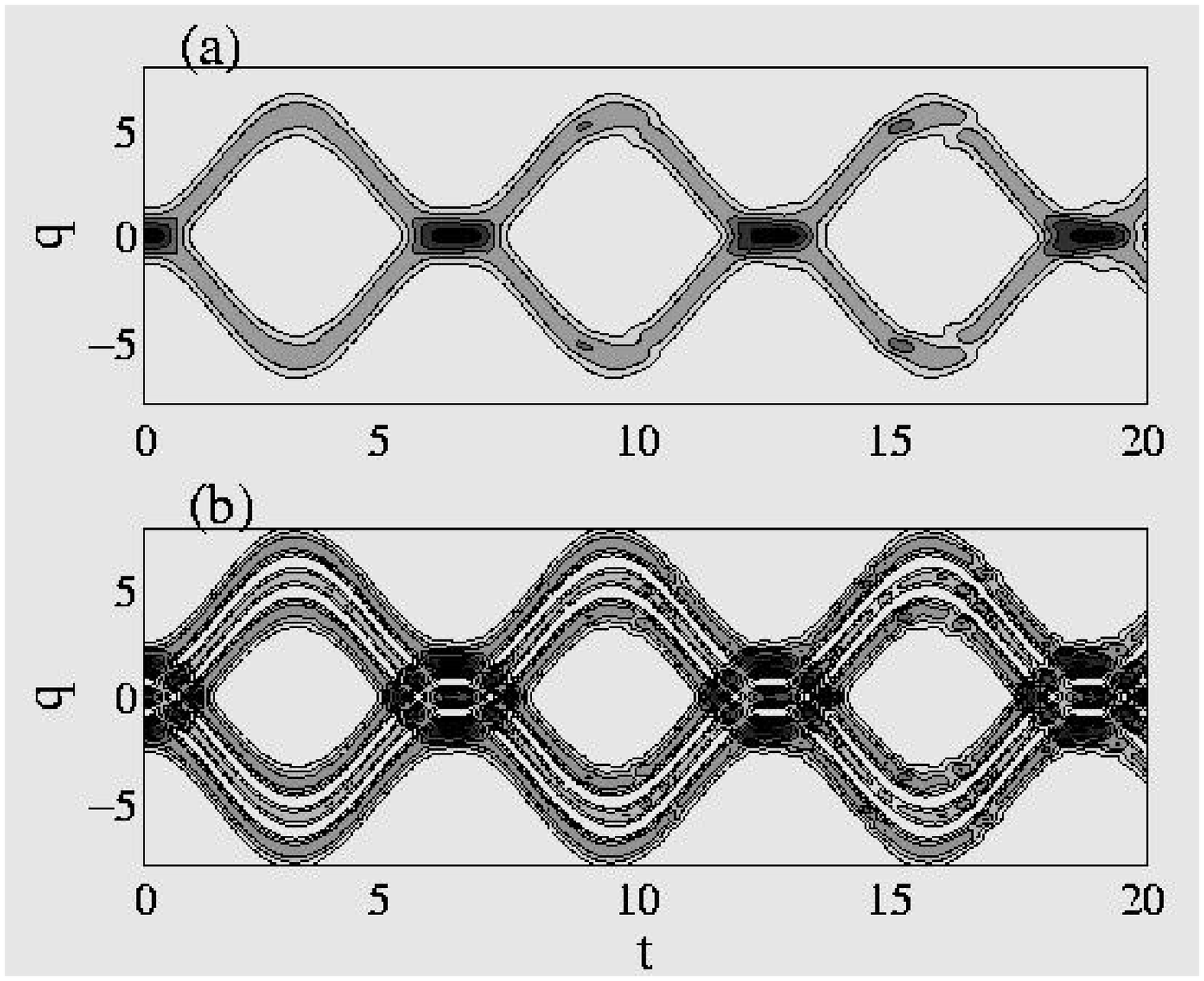}
\includegraphics[width=8cm]{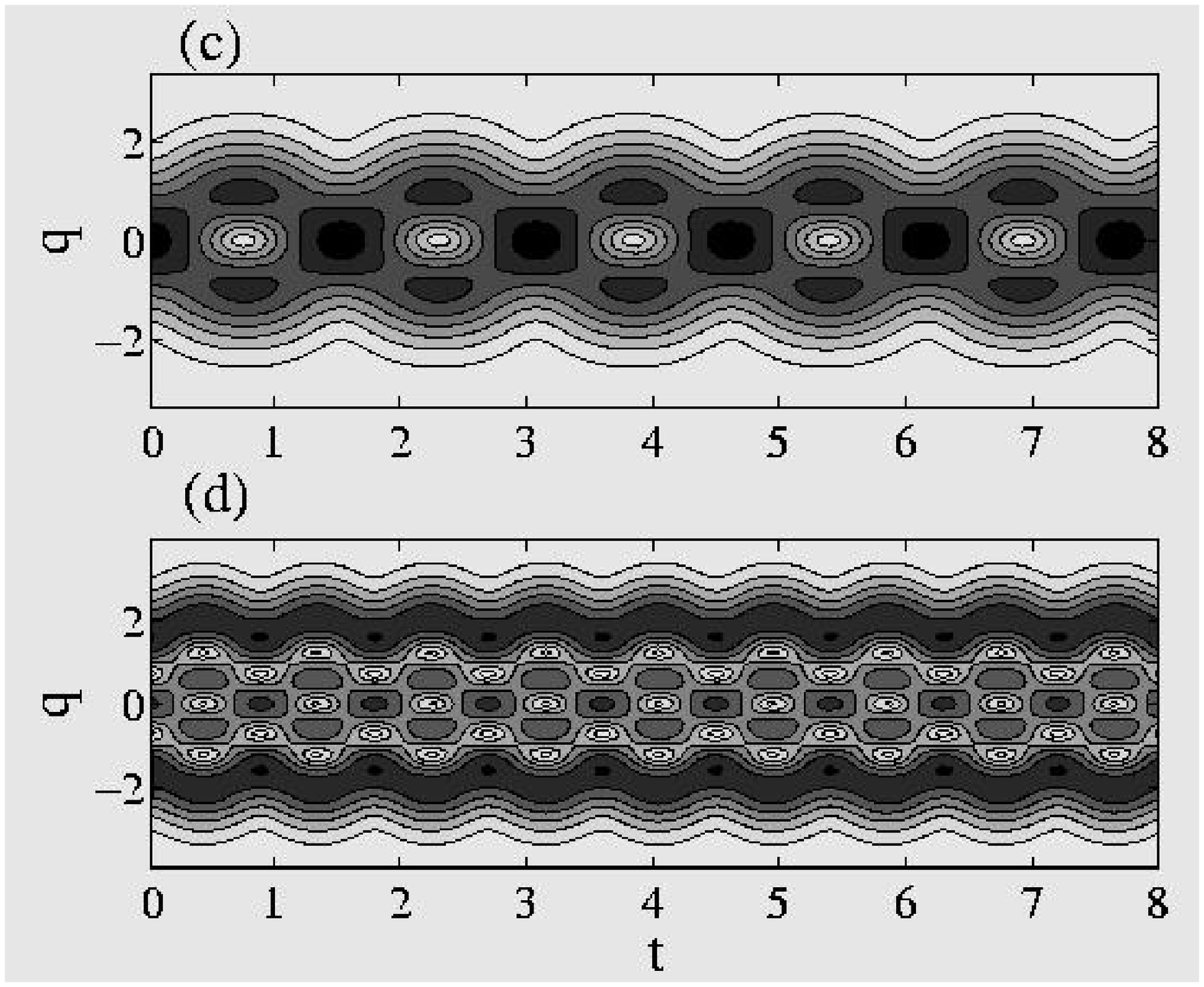}
\caption{\label{fig5} The absolute amplitudes of the
wave packets corresponding to the previous fig. \ref{fig4} (a)
without RWA (a) and (b) and with RWA (c) and (d). In (a) and (c) the
number of photons of the initial state is as in fig. \ref{fig5} (a);
$n=0$, while in (b) and (d) $n=2$. }
\end{center}
\end{figure}

Rabi oscillations \cite{rabi} describes the periodic population
swapping between some few internal levels, driven by an external
"field". For the JC model, Rabi oscillations manifest themselves by
population transfer between the upper and lower atomic state,
equivalent to an energy transfer between the field and the atom.
These oscillations are "exact" when the field is initially in a Fock
state and the systems are in resonance. Rabi oscillations have been
measured in several JC type of systems \cite{jcrabi}, and also
theoretically studied for the Rabi model \cite{rabirabi}. When the
non energy conserving terms are included, the structure of the
oscillations may be greatly affected. Figure \ref{fig4} gives two
examples where circles correspond to the Rabi model and solid line
to the JC one. In (a) we have $\Omega<\omega,g_0$, while in (b)
$\Omega>\omega,g_0$. The peaks of $\langle\sigma_z\rangle$ in fig.
\ref{fig4} (a) come with a period of $2\pi$ (in scaled units, where
$\omega=1$). The adiabatic frame is no longer appropriate for giving
an intuition of the behaviour, but instead one may consider the wave
packets as evolving on the two displaced oscillators. For the Rabi
model, the wave packet starts at the origin and then splits up into
two parts, respectively propagating down the two displaced
oscillators. After one classical period of oscillation the two
packets return to their original positions and the interference
gives rise to the peak. In the JC model, the wave packet will not
split into two disconnected parts because of the $p$ dependent
coupling which tend to bound the wave packet to the origin.
Therefor, there is no collapse in the Rabi oscillations. In (b), the
situation is adiabatic and since the adiabatic energy curves, fig.
\ref{fig1} (a), do not form two displaced oscillators, a collapse of
the Rabi oscillations is not as clear as in (a) for the Rabi model.
In fig. \ref{fig5} (a) and (c) we show the amplitudes of the wave
packets $\psi_\pm(q,t)$ corresponding to fig. \ref{fig4} (a), while
in fig. \ref{fig5} (b) and (d) we display the same but for an
initial Fock state with $n=2$ photons. The Rabi model wave packets
are displayed in (a) and (b), while (c) and (d) show the wave
packets for the regular JC model. Note that in (d) the wave packet
either has two or three peaks depending on if the field contains 1
or 2 photons respectively. The imperfect revivals come from the weak
coupling between the two oscillators, which induces the level
splitting of the oscillators at $q=0$.

\begin{figure}[ht]
\begin{center}
\includegraphics[width=8cm]{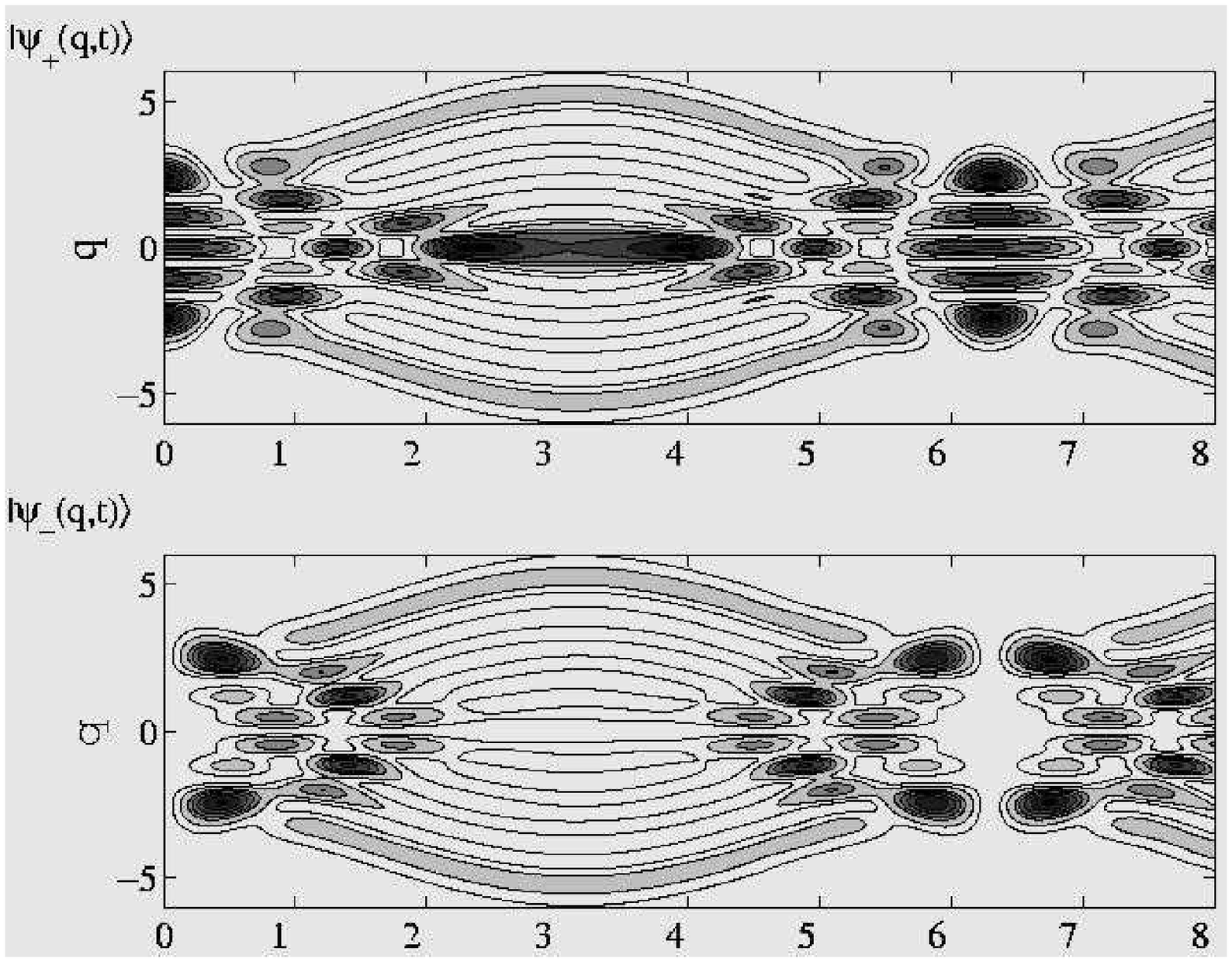}
\includegraphics[width=8cm]{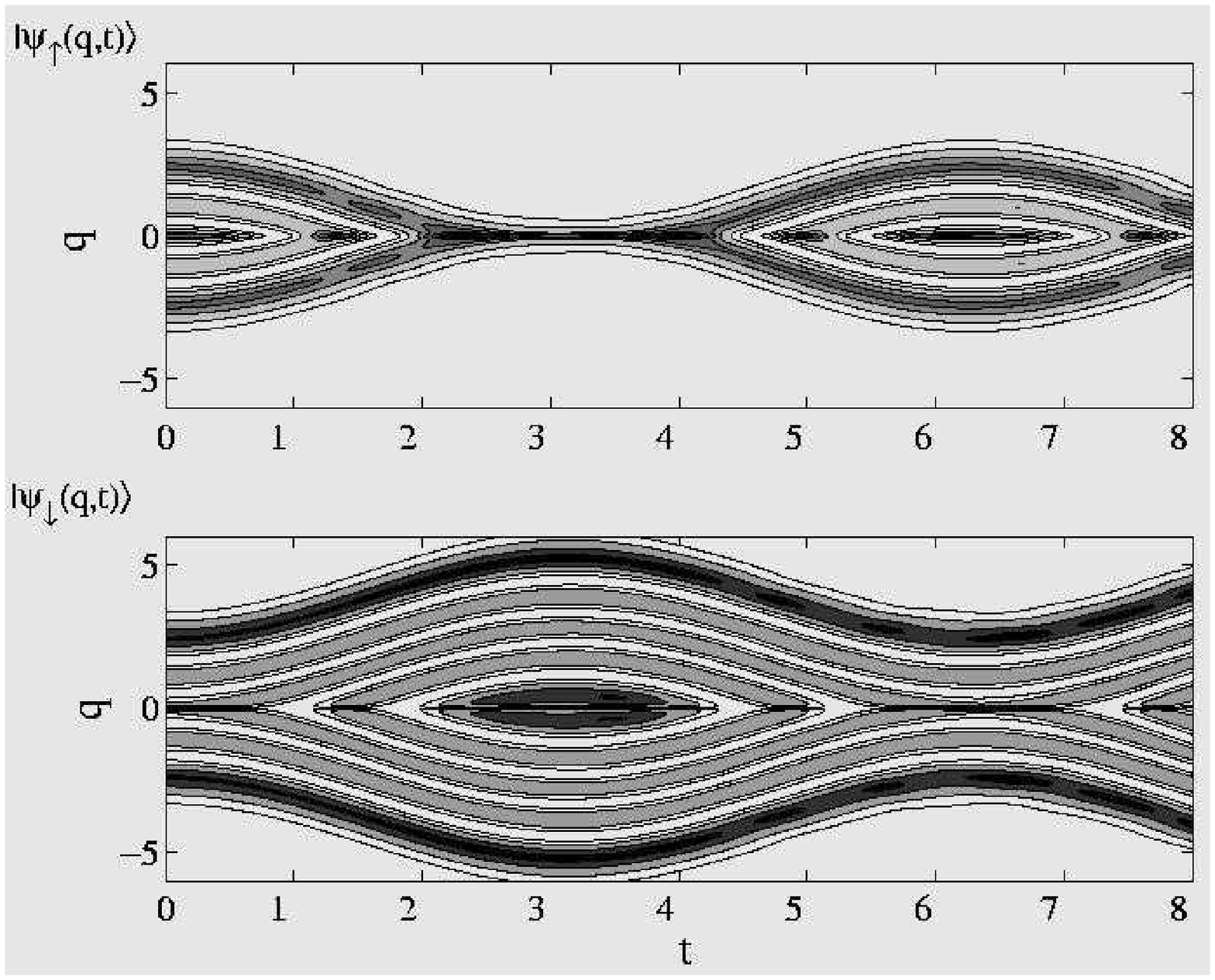}
\caption{\label{fig5b} The four different absolute amplitudes for the Rabi model in the strong coupling
regime corresponding to fig. \ref{fig1} (a), $g_0=1$ and
$\Omega=0.1$.  }
\end{center}
\end{figure}

Figure \ref{fig5b} gives the effect of the $p$ dependent coupling of
the two displaced oscillators in the JC model. It shows the
amplitudes $|\psi_\pm(q,t)|$ (upper plots) and
$|\psi_{\uparrow,\downarrow}(q,t)|$ (lower plots), for the Rabi
model with initial Fock state $|n=4\rangle$. The parameters are as
in fig. \ref{fig1} (d) and it is clear how the lower adiabatic wave
packet slides down the displaced oscillators, while the upper
adiabatic wave packet is squeezed at first. In the JC model, the
wave packets would look very similar to those of fig. \ref{fig5} (c)
and (d), since we know that the population Rabi oscillates between
the two Fock states $|n=4\rangle$ and $|n=5\rangle$. We can thus
conclude that the spreading and squeezing of the wave packets in the
Rabi model is due to the absence of the $p$ dependent coupling. This
has been verified numerically by artificially neglecting the $p$
term from the JC model and a similar evolution as in the Rabi case
is obtained. This effect of the $p$ term obviously holds for any
initial Fock state, also for large photon numbers $n$, which has a
broad wave packet. This may seem counterintuitive since for large
$n$ the wave packet is non vanishing where the two energy curves are
far apart and no coupling is expected to take part. However, as $n$
increases, the wave packet width also increases in momentum space
and the $p$-coupling between the curves becomes stronger.

\subsection{Collapse-revivals}
One of the more interesting phenomenon of the JC model is the
collapse-revival effect \cite{cr}. This is a direct consequence of
the "graininess" of the field: Various parts of the state will Rabi
oscillate with different frequencies, depending on the photon number
$n$, leading to a collapse of various physical quantities, such as the atomic inversion $\langle\sigma_z\rangle$. However,
after some particular time the parts may come back in phase and a
revival occurs. Revivals for evolving wave packets exist due to the anharmonicity of the
potential surface, and depending on the order $n$ of the anharmonicity terms ($\sim x^n$) different types of revivals are obtained, see the review \cite{wprevivals}. For initial squeezed sates, the JC model may also exhibit {\it super} and {\it fractional} revivals apart from the regular revivals \cite{fracrev}.

\begin{figure}[ht]
\begin{center}
\includegraphics[width=8cm]{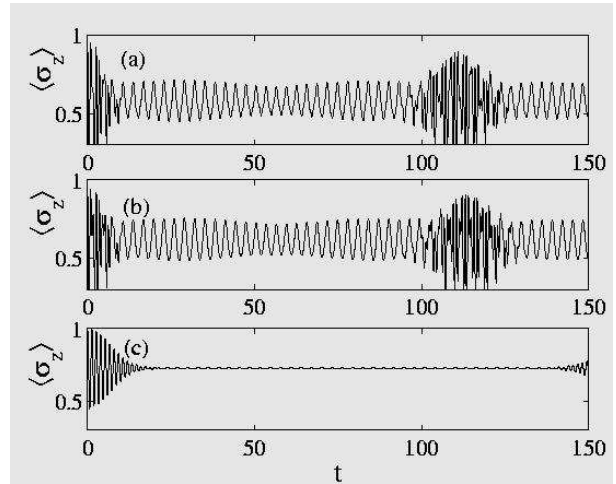}
\caption{\label{fig6} The atomic inversion for the Rabi model (a),
adiabatic approximation of the Rabi model (b) and JC model (c). Clearly
the JC model does not describe the dynamics properly in this case.
The parameters are $\Omega=5$, $g_0=0.3$ and $\nu=4$.}
\end{center}
\end{figure}

In \cite{jcrwa} it was shown, in the weak coupling limit, that the
inclusion of the counter rotating terms to the dynamics induces a
fast oscillating pattern on top of the regular JC one. This is,
however, in the weak coupling limit where we expect only small
changes, while we expect larger effects if the coupling is
increased. In the papers \cite{jcrwa,displace2} an approximate
method was introduced to study the collapse-revivals in the Rabi
model. In fig. \ref{fig6} the atomic inversion is plotted; (a) the
Rabi model, (b) the adiabatic approximation and (c) the JC model.
The parameters are in this case $\nu=4$, $\Omega=5$ and $g_0=0.3$.
Even though the fidelity (\ref{fidel}) has a minimum as low as
$\approx0.8$, the agreement between the exact (a) and adiabatic
approximated (b) results are very good. There is, however, a great
difference between the Rabi and JC results. The "plateaus" between
revivals are not as pronounced in the Rabi model, and the revivals
occur earlier.

\begin{figure}[ht]
\begin{center}
\includegraphics[width=8cm]{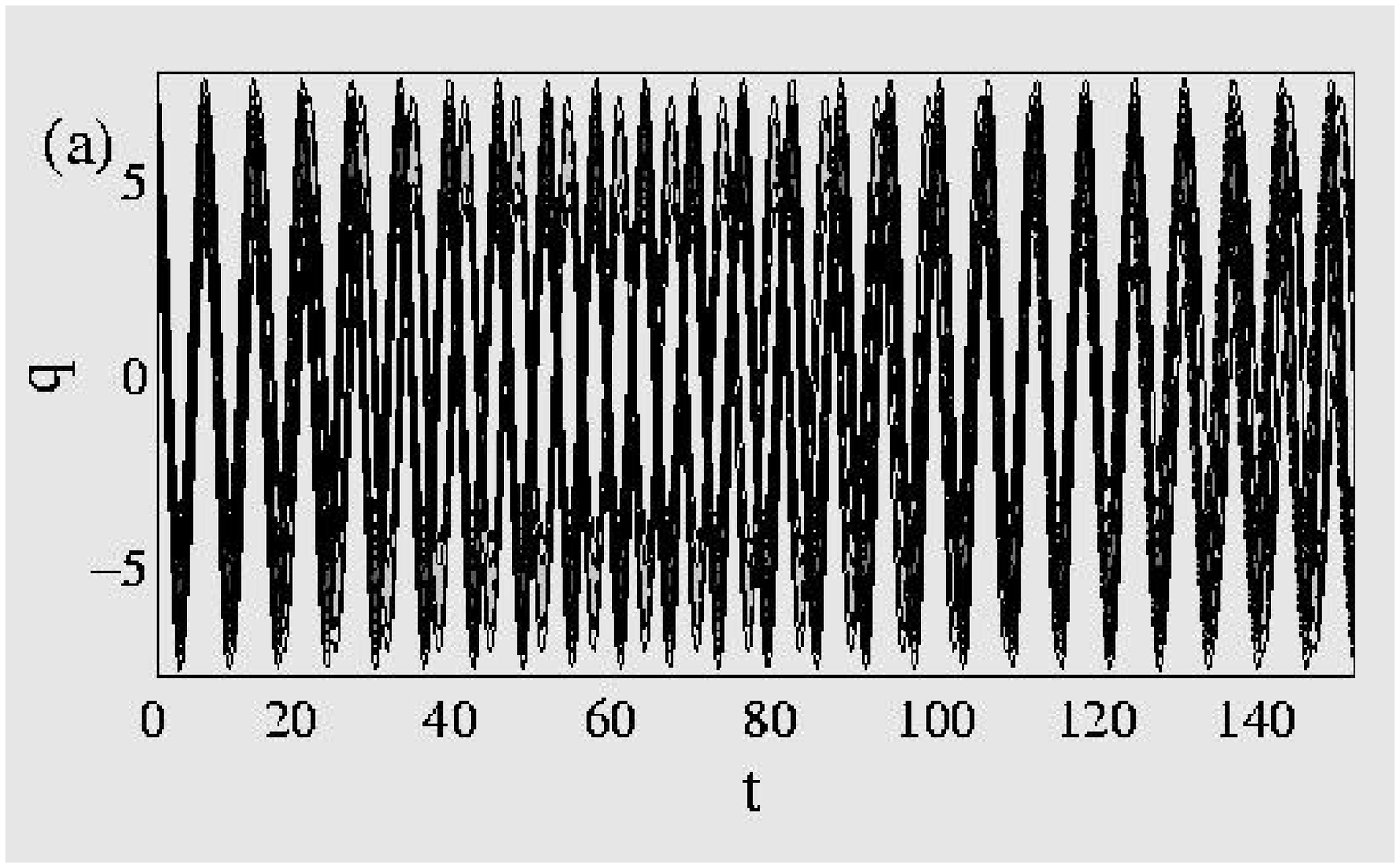}
\includegraphics[width=8cm]{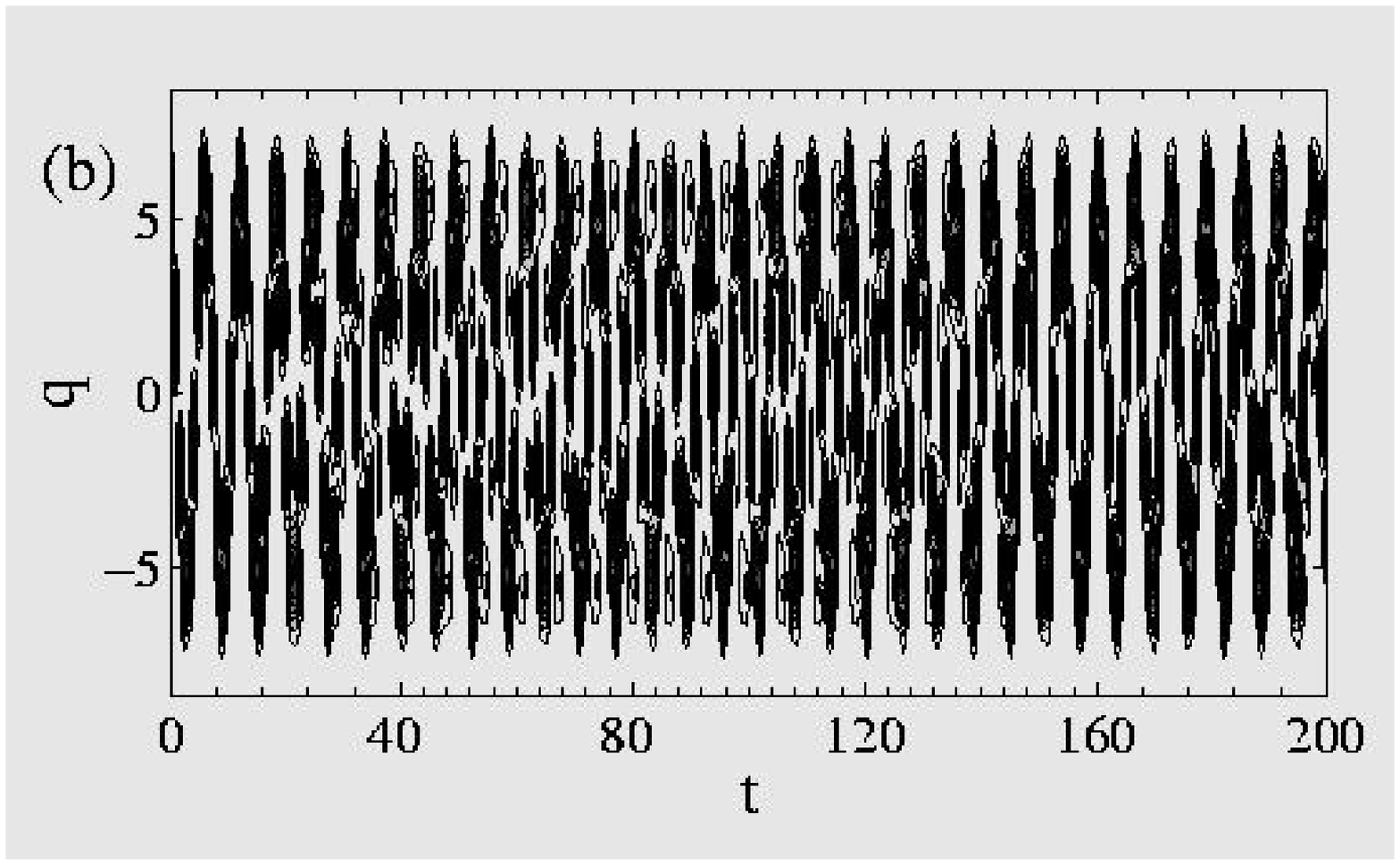}
\caption{\label{fig7} The absolute amplitudes of the wave
packets are shown in (a) corresponding to the Rabi case of the previous
figure \ref{fig6}, and (b) for the JC model. The two parts first come out of phase (collapse) and
then return back in phase (revival). It is seen that the revival
occurs at later times with the JC, as seen also in the previous
figure. }
\end{center}
\end{figure}

In the adiabatic regime the revivals may be easily understood and
one may estimate the revival time. The two adiabatic potentials
(\ref{adpot}), for example in fig. \ref{fig1} (a), are almost
harmonic with frequencies slightly bigger and smaller than $\omega$.
As the initial coherent states evolve on these potentials they will
first come out of phase and later return, giving the revival. The
anharmonicity and the non adiabatic contributions lead to non
perfect revivals. These arguments are confirmed in fig. \ref{fig7},
where the amplitudes of the wave packets for the Rabi model (a), and
for the JC model in (b), both corresponding to fig. \ref{fig6}, are
shown. A rough estimation of the revival time can be derived. In the
adiabatic limit we may consider the wave packets as evolving on the
disconnected adiabatic energy curves
\begin{equation}
\displaystyle{V_\pm^{Rabi}=\frac{x^2}{2}\pm\sqrt{\frac{\Omega^2}{4}+2g_0^2x^2}\approx\frac{\omega_{\pm,Rabi}^2x^2}{2}\pm\frac{\Omega}{2}}
\\ \\
\end{equation}
and working with the boson operators for the JC model
\begin{equation}
H_{JC}\approx\omega a^\dagger
a\pm\sqrt{\frac{\Omega^2}{4}+g_0^2a^\dagger
a}\approx\omega_{\pm,JC}a^\dagger a\pm\frac{\Omega}{2},
\end{equation}
where we have expanded the square roots and introduced the
renormalized frequencies
\begin{equation}
\displaystyle{\omega_\pm\equiv\omega_{\pm,Rabi}=\omega_{\pm,JC}=\sqrt{1\mp\frac{2g_0^2}{\Omega}}}\approx1\mp\frac{g_0^2}{\Omega},
\end{equation}
The revival times then approximate
\begin{equation}
\displaystyle{t_R=\frac{2\pi}{\omega_{-}-\omega_{+}}}\approx\frac{\pi\Omega}{g_0^2}.
\\ \\
\end{equation}
We see that this rough estimation gives identical revival times for
both the JC and the Rabi models and it is independent of the initial
state. Using the parameters of fig. \ref{fig6} we find
$t_R\approx174$. In the Rabi case, instead of an analytically
approximate result of the revival times given above, we numerically
calculate the curvatures of the adiabatic potentials. Such an
approach for the above example gives a revival time $t_R=125.8$,
which is much better compared to fig. \ref{fig6}. However, it should
be mentioned that measuring the curvature is not absolute since it
depends on the $x$-interval used. It is known that the revival time
depends on the initial state of the field. This is easily
understood, since the coupled harmonic oscillators depend on
"momentum" $p$, which is directly related to the amplitude of the
initial coherent state. For large photon numbers, the adiabatic
limit $g_0\sqrt{n}\ll\Delta$ breaks down, and the revival time given
above is of dubious validity. We can conclude that the above
analytical estimate is only valid in the very adiabatic limit, and
as such, it is interesting; in the adiabatic limit the revival time
does not depend on the initial state and consequently the average
number of photons $\bar{n}$.

\begin{figure}[ht]
\begin{center}
\includegraphics[width=8cm]{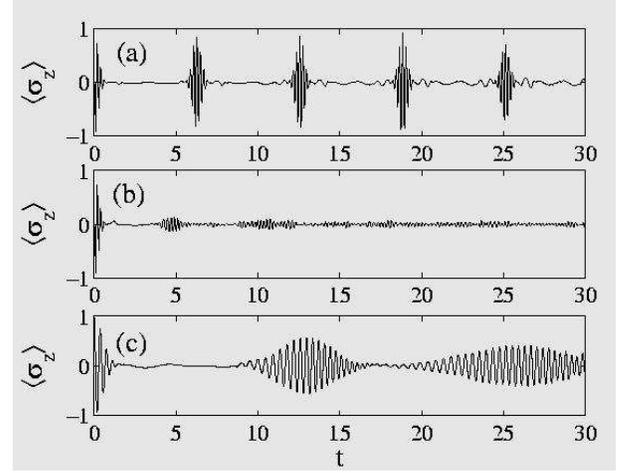}
\caption{\label{fig8} Same as fig. \ref{fig6}, but with $\Omega=0.2$
and $g_0=2$ instead. The revivals are most pronounced for the
non-RWA case (a).}
\end{center}
\end{figure}

\begin{figure}[ht]
\begin{center}
\includegraphics[width=8cm]{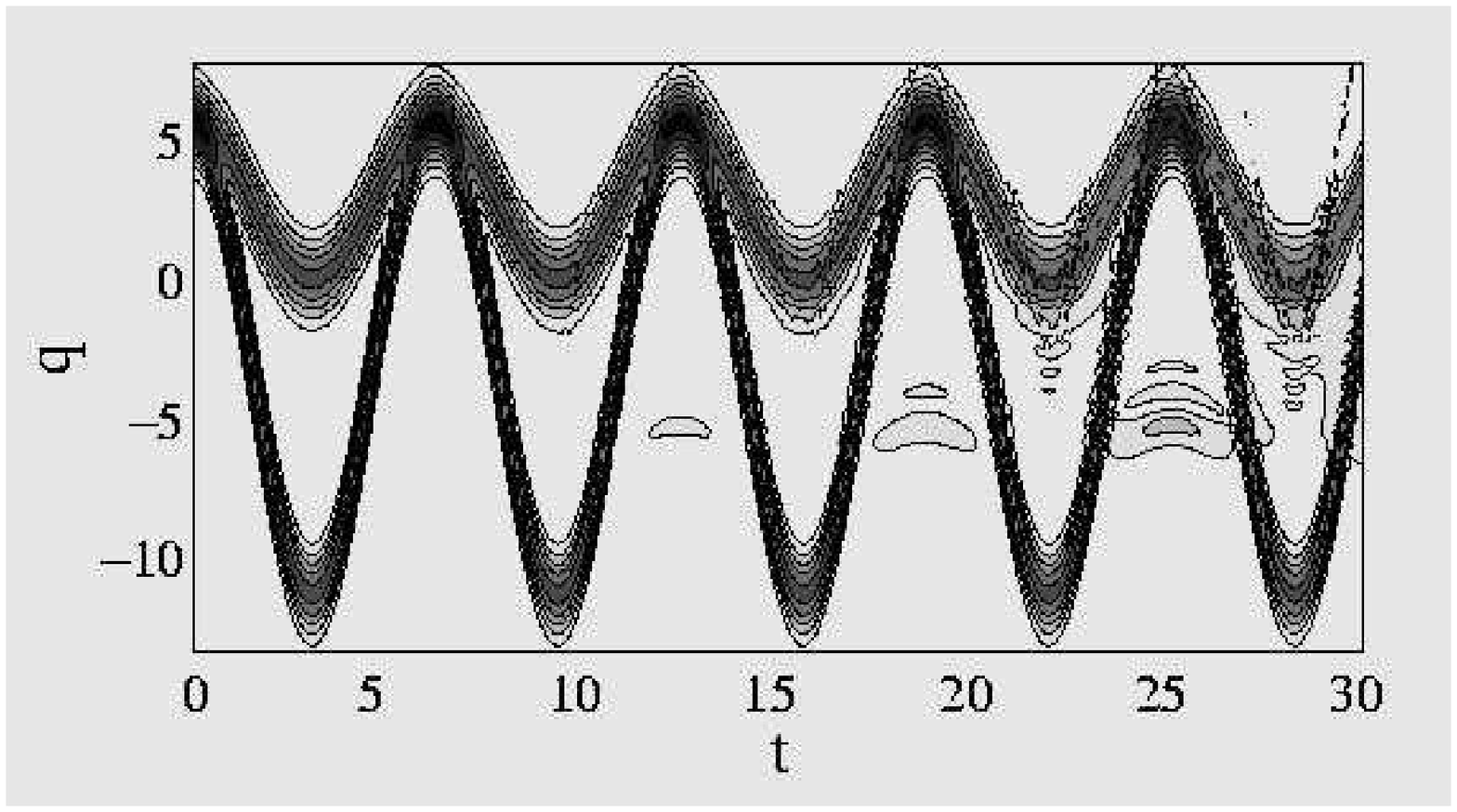}
\includegraphics[width=8cm]{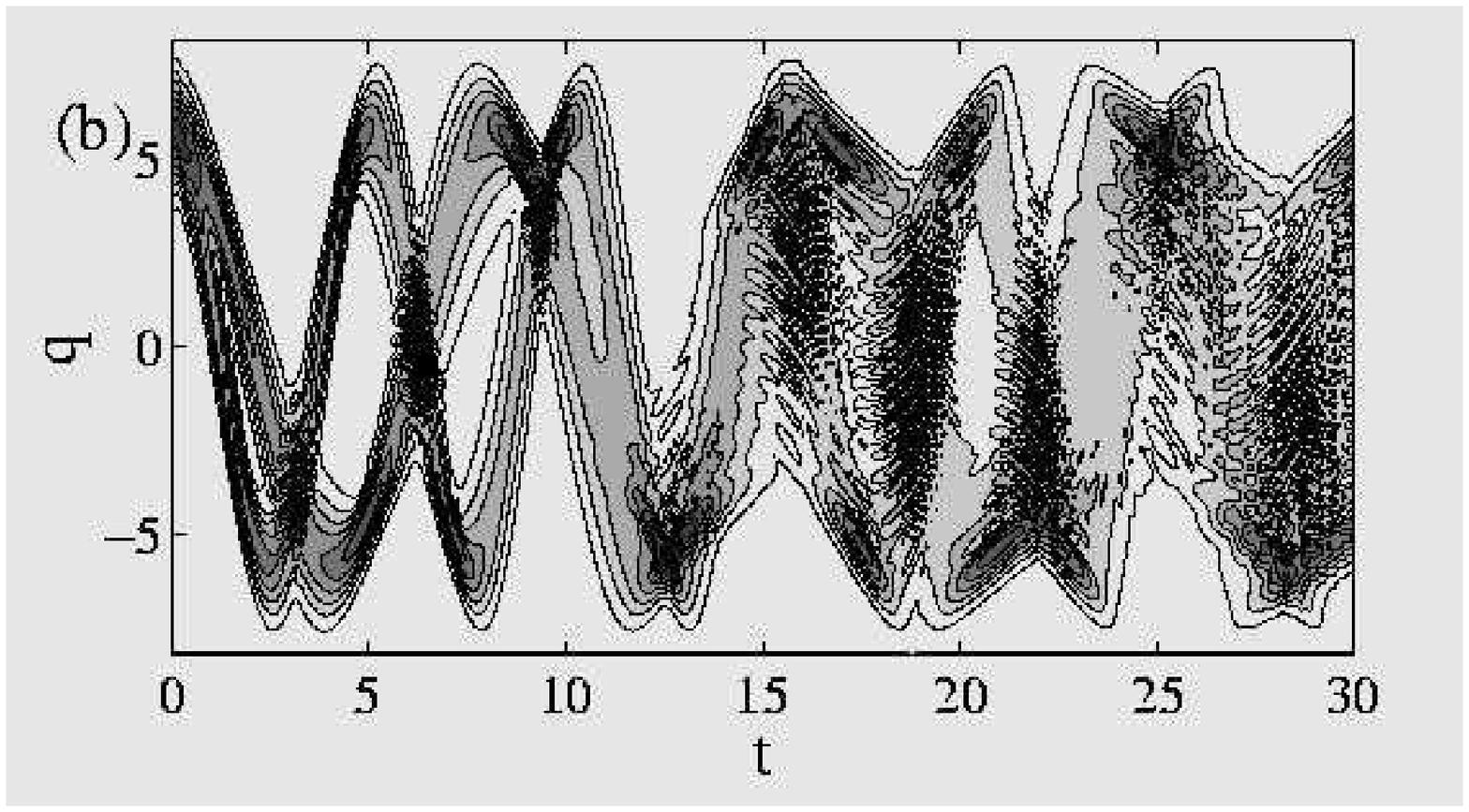}
\caption{\label{fig9} The absolute amplitudes of the original basis
($\{|\pm\rangle\}$) wave packets $\psi_\pm(q,t)$ corresponding to
the previous fig. \ref{fig8}. Plot (a) gives the amplitudes for the
Rabi model and the (b) the same for the JC model. }
\end{center}
\end{figure}

At some point, the adiabatic approximation breaks down and the above
picture of the collapse-revivals fails. Still one may have
interesting collapse-revival patters, as is seen in fig. \ref{fig8}.
This figure shows the same as fig. \ref{fig6}, but for the
parameters $\Omega=0.2$, $\nu=4$ and $g_0=2$. The revivals are
present in the Rabi model and seem very stable, while in the
adiabatic and JC situations they die out fast. The amplitudes of the
wave packets are displayed in fig. \ref{fig9}, where (a) gives the
Rabi case and (b) the JC situation. The plot clearly indicates the
existence of the collapse-revival effect. As the adiabatic
approximation is not valid we can no longer try to understand the
evolution from the an adiabatic point of view. However, since
$\Omega<g_0\langle n\rangle$ one may see the wave packet as evolving
along the diabatic energies described by the displaced harmonic
oscillators. Now they share the same frequency $\omega$ (the
anharmonicity comes from the avoided level crossing), so that the
oscillation periods are identical. Just like in fig. \ref{fig5} (a),
we see revivals in fig. \ref{fig8} (a), which occur when the
splitted wave packets return to their initial position. For other
parameter regimes, where neither the adiabatic nor the diabatic
frame holds,
 the inversion pattern for the Rabi model looks more random.

\subsection{Squeezing}
Another non-classical property of the field interacting with an atom
is squeezing, see \cite{jcsqueez}. The field is said to be squeezed
if one of the quadrature measures $\langle\Delta q^2\rangle=\langle
q^2\rangle-\langle q\rangle^2$ or $\langle\Delta p^2\rangle=\langle
p^2\rangle-\langle p\rangle^2$ is smaller than 1/2. The Squeezing in
the Rabi model has been studied in \cite{rabisqueez}, and it was
found that squeezing for the vacuum is mostly pronounced when
$\Omega\gg g_0$ and $g_0^2\gg\Omega$. The second condition implies
that the lower adiabatic energy curve has two minima, as in fig.
\ref{fig1} (b)-(d). Dipole squeezing for the Rabi model was
considered in \cite{disq}. In the standard JC model, the amount of
squeezing is improved for large amplitude initial states, meaning
states with a large average number of photons. The asymptotic
solutions in these limits have been applied to study squeezing of
the JC model \cite{gbsqueez}.

\begin{figure}[ht]
\begin{center}
\includegraphics[width=6cm]{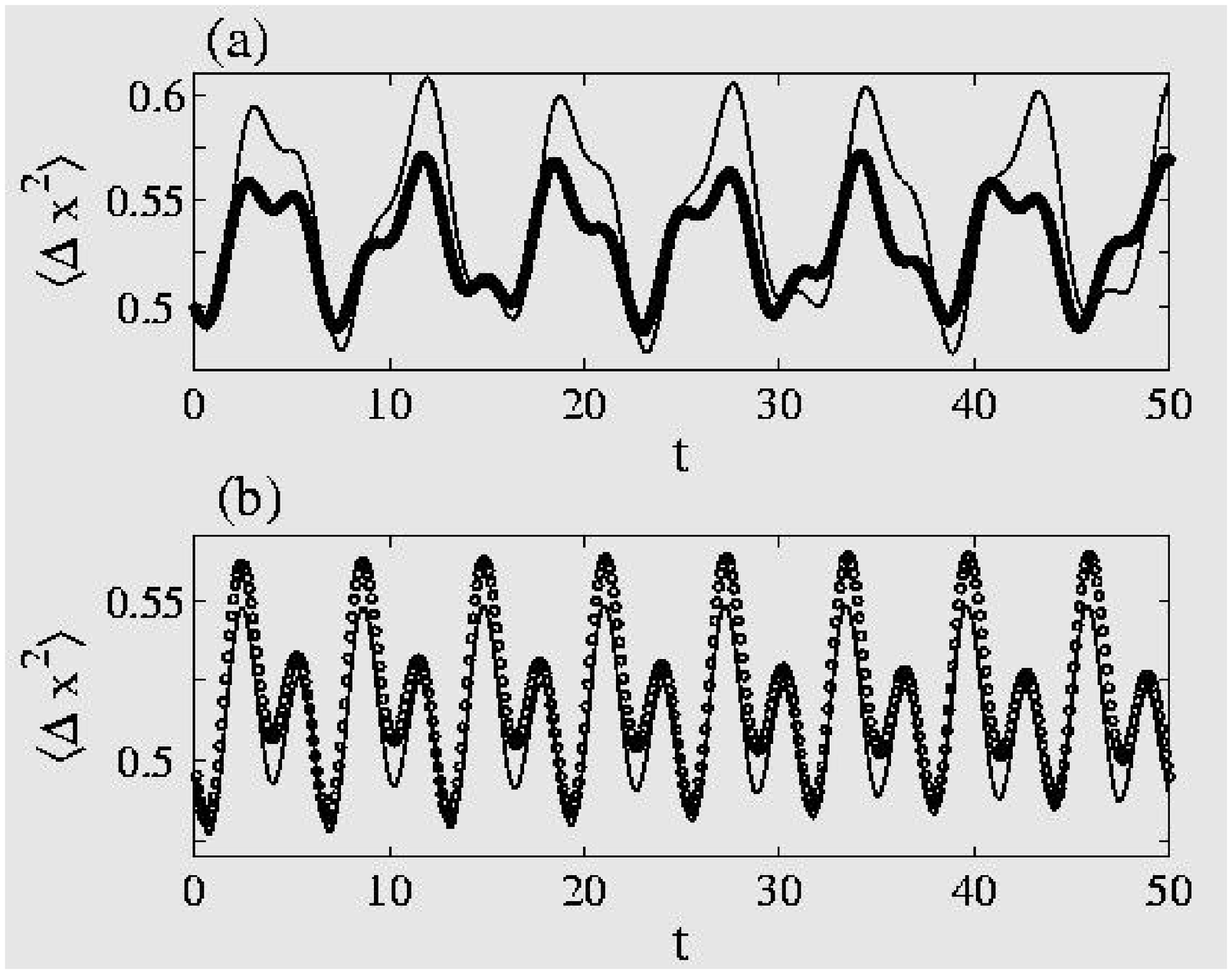}
\includegraphics[width=6cm]{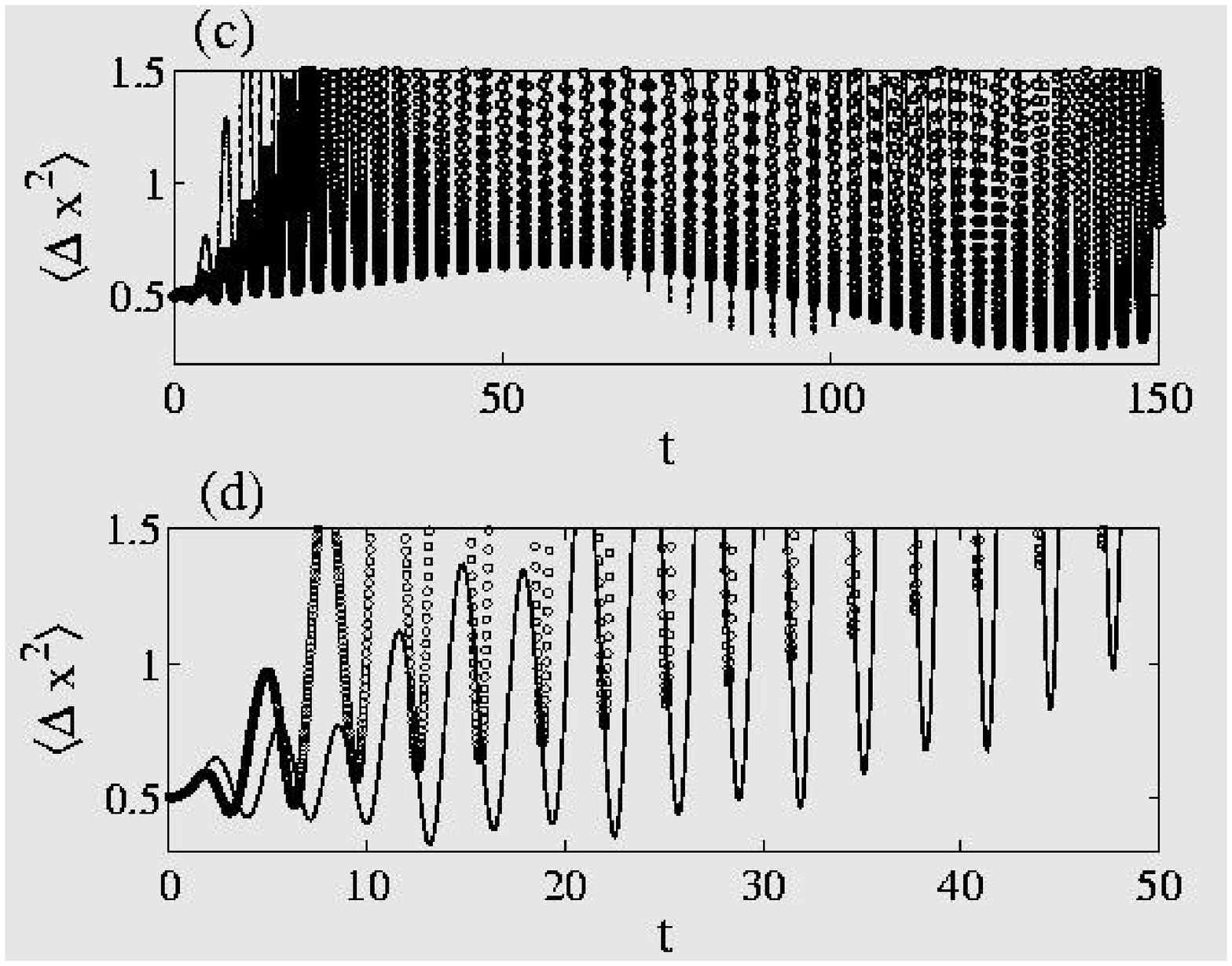}
\caption{\label{fig10} The absolute quadrature $\langle\Delta
x^2\rangle$ for the JC (circles) and the Rabi (solid) models. In (a)
$g_0=0.1$, $\Omega=0.2$ and $\nu=0$, in (b) $g_0=0.1$, $\Omega=2$
and $\nu=0$, in (c) $g_0=0.3$, $\Omega=5$ and $\nu=4$ and in (d)
$g_0=0.2$, $\Omega=1$ and $\nu=4$.   }
\end{center}
\end{figure}

In fig. \ref{fig10} the quadrature for $q$ is shown for four
different cases. In the first two plots, (a) and (b), the state of
the field is initially in vacuum and (a) $g_0=0.1$ and $\Omega=0.2$,
while in (b) $g_0=0.1$ and $\Omega=2$. In \cite{rabisqueez}, it is
shown that enhanced squeezing may be achieved if the atom is
initially in a linear combination of its internal states, such that
the wave packet is evolving mainly on one of the adiabatic energy
curves. This comes from the fact that squeezing comes from the anharmonicity of the energy
curves. Note that in both (a) and (b) the system is highly detuned
($\Omega=2,0.2$) resulting in a small squeezing. In (c) and (d) the
state of the field is initially in a coherent state with $\nu=4$,
and (c) $g_0=0.3$ and $\Omega=5$, and in (d) $g_0=0.2$ and
$\Omega=1$. Thus, in (d) the system is in resonance,
$\Omega=\omega$.

\begin{figure}[ht]
\begin{center}
\includegraphics[width=7.5cm]{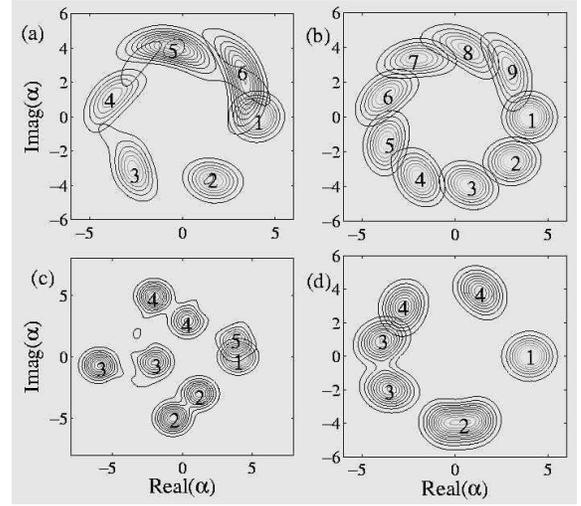}
\caption{\label{fig11} Evolution of the $Q$-function for the Rabi
model (left plots) and the JC model (right plots). In (a) and (b)
$g_0=0.3$, $\Omega=5$ and $\nu=4$, while in (c) and (d) we display a
resonant situation $\Omega=1$, and with $g_0=1$. Numbers indicate
time ordering and the final times (indicated by up to 9 in (a) and
(b) and by up to 5 in (c) and (d)) are $t_f=400$ for the upper plots
and $t_f=6$ for the lower ones. }
\end{center}
\end{figure}

\begin{figure}[ht]
\begin{center}
\includegraphics[width=6cm]{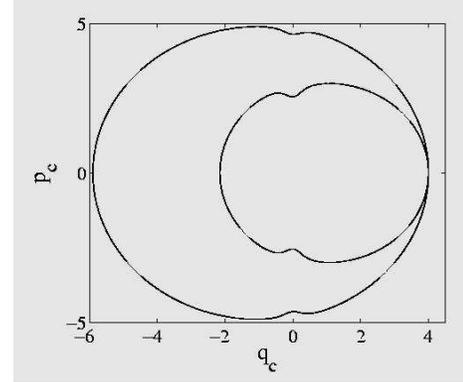}
\caption{\label{fig12} The classically obtained phase space trajectories (\ref{cpst}) corresponding to the previous example of fig. \ref{fig11} (c).   }
\end{center}
\end{figure}
Squeezing also manifests itself in the various quasi phase
distributions of the field. Here we give examples of the
$Q$-function \cite{lamb}, defined as
$Q(\alpha)=\pi^{-1}\langle\alpha|\rho_f|\alpha\rangle$, where
$|\alpha\rangle$ is a coherent state, and $\rho_f=\langle
+|\rho|+\rangle+\langle -|\rho|-\rangle$ is the field density
operator and $\rho$ the full atom-field density operator. For a
coherent state of the field, the $Q$-function is given by a
symmetric Gaussian in the complex
 $\alpha$-plane, while if it is squeezed, the Gaussian is no longer symmetric, but
squeezed in some direction. The $Q$-function for various types of JC
models has been studied in \cite{jcq} and for the Rabi model in
\cite{rabiq}. Figure \ref{fig11} gives the $Q$-function at different
times for the Rabi model (left plots) and for the JC model (right
plots), with an initial coherent state with $\nu=4$. In (a) and (b)
$g_0=0.3$ and $\Omega=5$, and therefor the system is largely detuned
which is seen since the $Q$-function does not split into two clear
parts, as in (c) and (d) where $g_0=1$ and $\Omega=1$. In the last
two plots the $Q$-function clearly splits up, however it is seen
that for the Rabi model the two parts of the function comes back
in phase contrary to what is seen for the JC model. The total time
evolved in (a) and (b) is $t_f=400$, while in (c) and (d) it is only
$t_f=6$. Since the coupling is larger in (c) and (d) together with
having the resonance condition, the characteristic timescales become
much shorter. During the splitting the system quantities collapse
and the revival occurs when the parts come back together again.
Therefor, in the Rabi model a revival takes place
very early, compared to the JC model. It should also be noted that
at half the revival time in the Rabi model the field has split into
a Schr\"odinger cat state \cite{cat} with approximately same phase but different
amplitudes, which was discussed by Phoenix \cite{rabiq}. This is not
the case of the JC model, where the splitting is solely in phase and not
amplitude.

\begin{figure}[ht]
\begin{center}
\includegraphics[width=8cm]{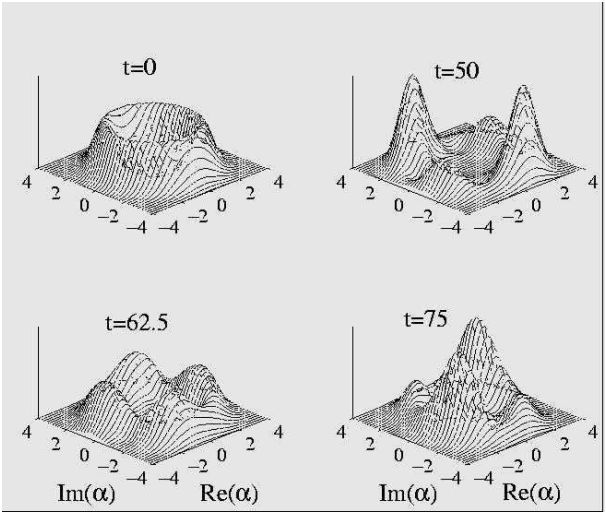}
\caption{\label{fig13} The evolution of a Fock state $|n=6\rangle$
in the Rabi model for different times and $\Omega=1$ and $g_0=0.5$.
}
\end{center}
\end{figure}

To understand this amplitude splitting we write
$\alpha=\frac{1}{\sqrt{2}}(q_c-ip_c)$ and approximate the real and
imaginary parts of $\alpha$ by its classical variables;
$\dot{q}_c=\frac{\partial H_\pm}{\partial p_c}$ and
$\dot{p}_c=-\frac{\partial H_\pm}{\partial q_c}$. In this
semi-classical approximation we use the adiabatic potential
(\ref{adpot}) as the classical potential governing the evolution
$H_\pm=\frac{p_c^2}{2}+V_\pm(q_c)$, and taking into account the
level splitting (replace $q_c$ by $|q_c|$ in the last term), we find
\begin{equation}\label{cpst}
 \begin{array}{l}
 \dot{q_c}=p_c, \\ \\
\displaystyle{\dot{p_c}=-q_c+\frac{16\Omega^2g_0^4q_c}{\left(\Omega^2+2g_0^2q_c^2\right)^3}\mp\frac{4g_0^2|q_c|}{\sqrt{\Omega^2+8g_0^2q_c^2}}},
\end{array}
\end{equation}
where the $\mp$-sign corresponds to which potential sheet
$V_\pm(q_c)$ is being used. In fig. \ref{fig12} we present the
classical phase space trajectories for the example of fig.
\ref{fig11} (c). The 'kinks' at $q_c=0$ are due to the second term
of $V_\pm(q)$ in eq. (\ref{adpot}), arising from the variations of
the potential. For the JC system, on the other hand, the Hamilton
equations of motion are of the form
\begin{equation}\label{cpst2}
 \begin{array}{l}
 \dot{q_c}=f_\pm(q_c,p_c), \\ \\
\displaystyle{\dot{p_c}=-f_\pm(p_c,q_c)},
\end{array}
\end{equation}
resulting in perfect circular trajectories due to the rotational
symmetry in 2-d. Another way to explain this splitting is given in
\cite{clsb}, where one considers the energy manifolds
\begin{equation}
\varepsilon_\pm(p,q)=\frac{p^2}{2}+\frac{q^2}{2}\pm\sqrt{\frac{\Omega^2}{4}+2g_0^2q^2}.
\end{equation}
The contours obtained by keeping
$\varepsilon_\pm(p,q)=\varepsilon_0$, where $\varepsilon_0$ is the
initial energy, govern the adiabatic trajectories in phase space.

As was noted above for the resonant situation and strong coupling,
$g_0\sim\omega$, the Rabi model gives very interesting behaviours,
different from the JC model. This does not only happen for initial
coherent field states. For a Fock state $|n\rangle$, the
$Q$-function is a "ring" with radii $|\alpha|=\sqrt{n}$ and thus
completely undetermined phase. In the JC model, a Fock state evolves
according to Rabi oscillations between $|n\rangle$ and $|n+1\rangle$
(the atom is initially in the excited state $|+\rangle$), and
therefor, the $Q$-function will have a "breathing" amplitude motion,
but the phase stays undetermined. For the Rabi model however, the
phase may not stay equally distributed over $2\pi$, and sort of cat
states may form. This is verified in fig. \ref{fig13}, where the
$Q$-function is plotted for four different times, $t=0,50,62.5$ and
75 for an initial Fock state with $n=6$ and $\Omega=1$ and
$g_0=0.5$.  Clearly, at some specific times $t$, the $Q$-function
consists of four distinct blobs and at other times two. For other
initial Fock states and other parameters one may obtain
$Q$-functions consisting of more than four blobs. These states could
be of interest for quantum information processing \cite{qinfo}.

\begin{figure}[ht]
\begin{center}
\includegraphics[width=6cm]{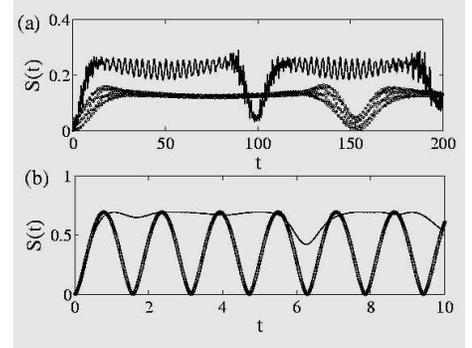}
\caption{\label{fig14} The entropy for an initial coherent state (a)
as in fig. \ref{fig6} but with $\nu=3$, and a Fock state (b) with
$g_0=1$, $\Omega=1$ and $n=0$. Circles display the entropy for the
JC model and solid line the Rabi results.
   }
\end{center}
\end{figure}

\subsection{Entanglement}

\begin{figure}[ht]
\begin{center}
\includegraphics[width=6cm]{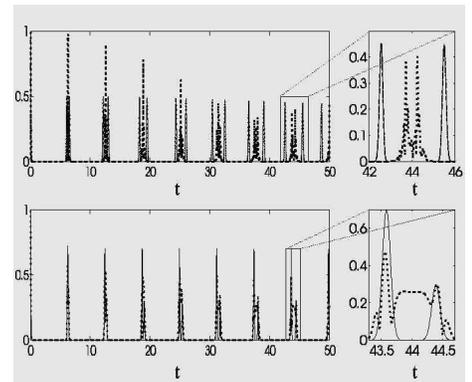}
\caption{\label{fig15} The absolute value of the autocorrelation
function for the JC (solid line) and the Rabi (dotted line) model,
for an initial coherent state with $\nu=15$ and $g_0=1$ and
$\Omega=1$ (a) and $g_0=0.3$ and $\Omega=5$ (b).  }
\end{center}
\end{figure}

The entanglement shared between the field and the atom is well
studied in the JC model \cite{entangle}. For example, it is known
that for initial coherent states of the field, the atom is entangled
with the field at all times except for half the revival times, where
it disentangles from the field provided a large field amplitude
\cite{gb}. Also for the Rabi model, entanglement has been analysed
\cite{rabient}. There are many measures of entanglement between two
interacting subsystems $A$ and $B$, where one of the more
established ones is the {\it von Neumann entropy}
\begin{equation}
S_A=-\mathrm{Tr}\left[\rho_A\log\left(\rho_A\right)\right],
\end{equation}
and $\rho_A=\mathrm{Tr}_B\left[\rho\right]$ is the reduced density
operator for subsystem $A$. For a pure disentangled state we have
$S_A=0$, and it can be shown that if the system starts out in a pure
state $S_A(t)=S_B(t)$ \cite{lieb}. For the two-level atom-field system
studied here, defining the coefficients
\begin{equation}
c_n^\pm(t)=\int\psi_n^*(q)\psi_\pm(q,t)\,dq,
\end{equation}
we find the eigenvalues $\lambda_\pm$ of the reduced density matrix
for the atom as
\begin{equation}
\begin{array}{lll}
\lambda_\pm & = &
\displaystyle{\frac{1}{2}\pm\sqrt{\frac{1}{4}+\left|\sum_nc_n^+(t)c_n^{*-}(t)\right|^2}}\\
\\& & \displaystyle{\sqrt{+\sum_n\left|c_n^+(t)\right|^2\left(\sum_n\left|c_n^+(t)\right|^2-1\right)}}
\end{array}
\end{equation}
and the entropy
\begin{equation}
S_A=-\lambda_+\log\left(\lambda_+\right)-\lambda_-\log\left(\lambda_-\right).
\end{equation}
Other measures for entanglement between bipartite systems are
possible \cite{bengtsson}.

Examples of the entropy are presented in fig. \ref{fig14} for an
initial coherent state (a), corresponding to the example of fig
\ref{fig6} but with $\nu=3$, and a Fock state (b) at resonance
$\Omega=1$ and $g_0=1$. As expected, the entanglement is in general
larger in the Rabi model. For the coherent states, we see the
typical disentanglement at the revival time in both models. For a
Fock state we know that in the JC model, the atom disentangle from
the field periodically with the Rabi oscillations. This is not the
case in the Rabi model, since a set of Fock states of the field will
be coupled to the atom and the disentanglement must occur between
all field states and the atom simultaneously.

\subsection{Autocorrelation function}

\begin{figure}[ht]
\begin{center}
\includegraphics[width=6cm]{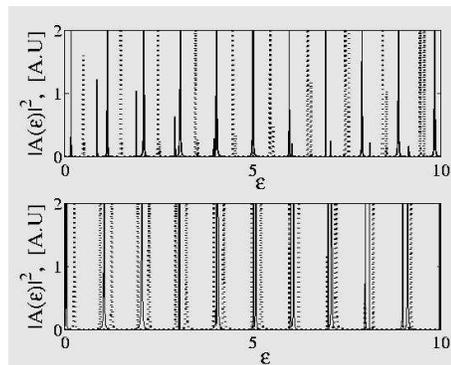}
\caption{\label{fig16} The absolute value squared of the fourier
transformed autocorrelation functions of the previous fig.
\ref{fig15}. Dotted line corresponds to the Rabi model, while solid
line the JC model. }
\end{center}
\end{figure}

The autocorrelation function
\begin{equation}
A(t,s)=\langle\Psi(t)|\Psi(s)\rangle,\hspace{2cm}t>s,
\end{equation}
contains much information about the dynamics of a given system. The
mathematical properties of the autocorrelation function have been
studied by numerous authors \cite{auto}. Here we will take time
$s=0$ (and for simplicity denote the autocorrelation function by
$A(t)$), and thus, the autocorrelation function measures fidelity
between the initial state and the time evolved state. Expanding the
initial state in eigenstates of the Hamiltonian (assuming a discrete
spectrum), $|\Psi_0\rangle=\sum_nc_n|\psi_n\rangle$, we get
\begin{equation}
A(t)=\sum_n\mathrm{e}^{iE_nt}|c_n|^2.
\end{equation}
Fourier transforming the autocorrelation function then gives a
direct relation to the spectrum, $E_n$ of the Hamiltonian
\begin{equation}
|A(\epsilon)|\propto\sum_n|c_n|^2\delta(\epsilon-E_n).
\end{equation}
Note that in the numerical simulations, time $t$
is given on a lattice and the lattice length $L_t$ and spacing $dt$
determine the $\epsilon$-lattice according to the FFT, and also the
widths of the $\delta$-function peaks in the spectrum.

In fig. \ref{fig15} we give an example of the autocorrelation
function $A(t)$ for an initial coherent state with amplitude
$\nu=15$, and parameters $g_0=1$ and $\Omega=1$ (a) and $g_0=0.3$
and $\Omega=5$ (b). The dotted line corresponds to the Rabi model
while solid line to the JC model. We note that the peaks occur at
around times $t=2n\pi$ ($\omega=1$), $n=0,1,2,...$, and that there
are two such reviving peaks for each $2n\pi$ and they become more
separated for larger times. The autocorrelation function also
contains the information about the revival times; when the distance
between the two seperated peaks becomes equal to $2\pi$,
interference of the wave packets take place resulting in a revival.
Since the separation is faster for the JC model in the (a) plot, the
revivals will occur earlier in the JC model compared to the Rabi
model. The Rabi model seperation is not as smooth as for the JC
model which will tend to smear out the revivals, which was already
seen in fig. \ref{fig6}. In (b), corresponding to fig. \ref{fig6},
the separation speed is slightly larger for the Rabi model and the
revival times is therefor shorter than for the JC model. Measuring
the revival times from fig. \ref{fig15} (a) and (b) for the JC model
one finds $t_R\approx 94.2$ and $t_R\approx343.4$ respectively,
which coincide well with the known analytical approximate expression
\cite{scully}
\begin{equation}
t_R\approx 2\pi\frac{\sqrt{\bar{n}}}{g_0}\left(1+\frac{\Delta^2}{4g_0^2\bar{n}}\right)^{1/2}.
\end{equation}
Since the wave packet evolves on potential surfaces that are not
fully harmonic, small spreading occur and the peaks in the
autocorrelation function will as well broaden. This in turn leads to
decreasing revival amplitudes and the characteristic revival widths
become longer, see fig. \ref{fig8} (c).

Figure \ref{fig16} shows one part of the fourier transformed
autocorrelation functions of fig. \ref{fig15}. Clear peaks
indicating the eigen enrgies are noted for both models. There seems
to be two different energy spacings between the peaks, one slightly
larger and one slightly lower than $\epsilon=1$. In the JC model
this is easily understood from the analytic expression
(\ref{jcsol2}) for the energies, which for large values on $n$ can
be linearly approximated over short $n$-intervals. From the fact
that $\bar{n}\gg1$ it follows that also $q\gg1$ and we are in the
adiabatic regime, where the potential curves are well described by
two centered harmonic oscillators with frequencies slightly larger
and smaller than one. This, of course, also applies to the Rabi
model, and in \cite{pertrabi} it is shown, in a perturbative way,
that the spectrum of the Rabi model approaches an equidistant one in
the strong coupling limit. When $\bar{n}$ is decreased the
linearization of the energies (\ref{jcsol2}) is no longer justified,
which as well follows from the fact that the adiabatic approximation
breaks down in such regimes.

\section{Conclusions}\label{conc}
In this paper we have used a wave packet approach to get a deeper
understanding of two of the most commonly used systems in quantum
optics and solid state physics; the JC and Rabi models. The wave
packet method is purely numerical, but once the evolved wave
packets are obtained, all various physical quantities are easily
obtainable. Also, for the JC model, which is analytically solvable,
one is often left with an infinite sum representing some physical
quantity, and either approximations or numerical methods are used to
get a "closed" form of the quantity. More important, from
the knowledge of how the wave packet evolves on the two coupled
potential curves, one can understand several phenomena such as Rabi
oscillations, entanglement and collapse-revivals, and even find
approximate results for some quantities in the adiabatic or diabatic
limits. Another advantage of this approach is that it is numerically
easily obtained for any parameter regimes, especially in the strong
coupling regime, where other approaches such as truncation methods
tend to be cumbersome.

The focus has been to study some of the known phenomena of the
models and compare the two. Several differences have been observed,
and, of course, especially when the RWA is known to break down. The
comparison between the two models is, however, mainly to understand
the wave packet evolutions, and not to point out the physical
differences between the two. For very strong couplings, the Rabi
model shows many interesting effects, for example splitting of the
phase-space function into several parts, and also splitting in
amplitude. This effect has not been studied in detail earlier
\cite{rabiq}. Another interesting aspect seen from the wave packet
simulations is the effect of the "momentum" coupling between the two
energy curves, present in the JC model. It has been shown that this
term bounds the wave packet close to the origin, making it not slide
down the two energy curves. The link between the JC model and curve
crossing problems is of great interest since knowledge from either
of the two may give insight into the other one. The semi-classical
treatment presented in subsection \ref{sslz} shows how to derive
time dependent two-level Hamiltonians, which have aspplications in
several fields of quantum mechanics \cite{ccross,ccross2}.

The method of wave packet propagation may easily be applied to
similar or extended models, for example Kerr systems \cite{kerr},
pumped systems \cite{pump}, the Dicke model \cite{lamb,dickem} and multi-level models \cite{multilev} and it is believed that much insight of these systems may be gained.
The relation between the models studied in this paper and models
describing diatomic molecular dynamics is planed for a future
project (see also \cite{molekyl}), and especially study the effect of anharmonicity in the
potential curves on the physical quantities. For such analysis it is likely that shape invariant potential techniques (SUSY) may be applicable \cite{susy}. Interference effects
between the evolved wave packets, but for a molecular system, will
be presented elsewhere \cite{asa}. These results may have
interesting applications in the JC or Rabi model.

\begin{acknowledgments}
I acknowledge Prof. Maciej Lewenstein and Prof. Stig Stenholm
for helpful discussions and the Swedish Goverment/Vetenskapsr\aa det
and EU-IP Programme SCALA (Contract No. 015714) for support.
\end{acknowledgments}

\end{document}